\documentclass[a4paper,11pt]{article}
\usepackage{graphicx}
\usepackage{jheppub}
\usepackage[T1]{fontenc} 
\title{\boldmath
Birth and death of entanglement between two accelerating Unruh-DeWitt detectors coupled with a scalar field} \author[a]{Dawei Wu}
\author[a]{Shan-Chang Tang}
\author[a,b,c,1]{Yu Shi \note{Corresponding author.}}
\affiliation[a]{ Department of Physics, Fudan University, Shanghai 200438, China}
\affiliation[b]{ School of Physical Sciences,  University of Science and Technology, Hefei 230026, China}
\affiliation[c]{ Shanghai Research Center for Quantum Science and CAS Center for Excellence in Quantum Information and Quantum Physics, University of Science and Technology of China, Shanghai 201315, China}

\emailAdd{16110190012@fudan.edu.cn}
\emailAdd{tangshanchang@sina.com}
\emailAdd{yushi@fudan.edu.cn}

\abstract{We consider two accelerating Unruh-DeWitt detectors  coupled linearly or quadratically with a scalar field. We show that entanglement can be created by acceleration, and is divergent only when the two detectors coincide. For linear coupling, entanglment decreases monotonically with the increase of  acceleration. For quadratic coupling, entanglement behaves non-monotonically.}

\begin{document}
\maketitle
\flushbottom

\section{Introduction}

Quantum field theory in curved spacetime, in which the curvedness of the spacetime is considered but the gravity is not quantized \cite{Wald-1995},  has made many remarkable predictions, such as the Unruh effect \cite{Unruh-1976} and Hawking radiation \cite{Hawking-1975}. An approach  to studying  these phenomena is to consider a particle detector, which provides a good illustration of the notion of  particle  in quantum field theory. As the simplest model, the Unruh-DeWitt detector\cite{DeWitt-1979, Unruh-1984}  describes a two-level system coupled linearly with a scalar field, similar to an atomic dipole coupled with an electromagnetic field \cite{Pozas-Kerstjens-2016}. It is natural to extend the consideration to other kinds of interaction, for instance, the $\hat{\Phi}^2$ coupling, the coupling with a Dirac field~\cite{Takagi-1986}, etc. However, these interactions run into severe divergence problems in  their correlation features \cite{Sachs-2017}, which reveals the entanglement of the quantum fields.

The Reeh-Schlieder theorem~\cite{Reeh-1961} shows that in quantum field theory, all field variables in any spacetime region  are entangled with those in the other regions,  thus the vacuum state is entangled, as  can be explicitly described by the thermofield double~\cite{Maldacena-2003}, which is closely related to the Unruh effect. Given that the field is entangled, people are interested in the so-called  entanglement harvesting~\cite{Salton-2015}, i.e. swapping  the correlation in the quantum field to that between particle detectors. There are a few different scenarios, in which  two detectors may be  static, uniformly accelerating, or freely moving in a gravitational field.  Because of the divergences mentioned above, however, few studies consider  particle detector models with nonlinear coupling with the quantum fields.

In this paper we focus on the entanglement harvesting of two uniformly accelerating detectors, both of which coupled quadratically with a scalar field. Other than working in the inertial frame, as in most of the previous studies, we tackle the problem from the Rindler observers' perspective, which is more natural for our purpose, and the divergences appear only when these two detectors conincide in spacetime. Furthermore, for  $\Phi^2$ coupling, there appears a striking feature that the entanglement does not decrease monotonically with the acceleration, in contrast to the usual situation that  the acceleration enhances entanglement harvesting. This feature was noticed only  recently~\cite{Liu-2021}, and is quite generic in our case, although  for large enough acceleration, the entanglement decreases in the end till  the sudden death~\cite{Yu-2009}. We will also discuss how the entanglement depends on the distance between the detectors.

This paper is organized as the following. In section 2 we will give a brief introduction to the Rindler modes, which are crucial in our calculation, and we will derive the reduced density matrix of our model which determines the entanglement properties. In section 3 and 4 we will study detectors that are linearly and quadratically coupled with the scalar field respectively. It can be seen that in these two different circumstances, the entanglement exhibits different features.  In section 5, we will summarize our results and give an outlook.

\section{Set up}

Before we introduce the model, it is worthwhile to briefly review  quantum field theory in the Rindler wedge since we will be working in the Rindler perspective.

The right Rindler coordinate $(\tau, \xi, x,y)$ is related to the Minkowski coordinate $(t,x,y,z)$ as
\begin{equation}
  t=a^{-1}e^{a\xi}\sinh a\tau,\,\,\, z=a^{-1}e^{a\xi}\cosh a\tau,
\end{equation}
while $(x,y)\equiv \mathbf{x}_\bot$ is the same as in the Rindler coordinate. A  massive scalar field can be expanded in the right Rindler wedge as~\cite{Crispino-2008}
\begin{equation}
  \hat{\Phi}(\tau,\xi,\mathbf{x}_\bot)=\int_0^\infty d\omega\iint d^2\mathbf{k}_\bot \left(\hat{a}^R_{\omega\mathbf{k}_\bot}v^R_{\omega\mathbf{k}_\bot}+\hat{a}^{R\dag}_{\omega\mathbf{k}_\bot}v^{R*}_{\omega\mathbf{k}_\bot}\right), \label{eqn: modeexpansion}
\end{equation}
where the lower bound of the integration over $\omega$  should be $0$,
\begin{equation}
  v^R_{\omega\mathbf{k}_\bot}=\left[\frac{\sinh(\pi\omega/a)}{4\pi^4a}\right]^{
  1/2}K_{i\omega/a}\left(\frac{\kappa}{a}e^{a\xi}\right)
  e^{i\mathbf{k}_\bot\cdot\mathbf{x}_\bot-i\omega\tau},
\end{equation}
with  $\kappa\equiv\sqrt{\mathbf{k}_\bot^2+m^2}$ and $K_\nu(x)$ is the modified Bessel function.

Since the annihilation operator for a certain mode $\hat{a}^R_{\omega\mathbf{k}_\bot}$ in the Rindler wedge is different from that in the Minkowski spacetime, the vacuum state of the scalar field is different in the two coordinate systems.

The Minkowski vacuum can be expressed in  the right  Rindler wedge as~\cite{Crispino-2008}
\begin{equation}
  |0_M\rangle=\prod_{\omega\mathbf{k}_\bot}\left(C_\omega\sum_{n_{\omega\mathbf{k}_\bot}=0}^\infty e^{-\pi n_{\omega\mathbf{k}_\bot}\omega/a}|n_{\omega\mathbf{k}_\bot},R\rangle\otimes|n_{\omega\mathbf{k}_\bot},L\rangle\right),\label{eqn: thermofield-double}
\end{equation}
where $C_\omega=\sqrt{1-\exp(-2\pi\omega/a)}$,  the product over $\omega$ and $\mathbf{k}_\bot$ should be understood as the limit of the discrete case. The mathematically rigorous formulation is based on the  S matrix  between the two Fock spaces~\cite{Wald-1995}.

Now we consider two Unruh-DeWitt detectors which accelerate in $+z$ direction with the same acceleration $a$, but are separated in the $x$ direction with distance $x_0$. The Hamiltonian in the interaction picture is
\begin{equation}
  H_I=H_A+H_B,
\end{equation}
where, with $j=A, B$,
\begin{equation}
  H_j=\lambda_j\chi_j(\tau)\int_{\Sigma_j} {\mathcal{O}}(\tau,\mathbf{x})\left[f_j(\mathbf{x})\sigma_j^+e^{i\Omega_j\tau}
  +f_j(\mathbf{x})^*\sigma_j^-e^{-i\Omega_j\tau}\right]\sqrt{-g}d^3\mathbf{x},
\end{equation}
where $\lambda_j$ is the coupling constant, $\chi_j$ is the switch  function, $ {\mathcal{O}}$ is the field operator, which is  $ {\Phi}$ or $ {\Phi}^2$,  $f(\mathbf{x})$ represents the shape of the detector; $\sigma_j^+$ and $\sigma_j^-$ are the raising and lowering operators of the two-level system, $\Omega_j$ is the energy gap of the system, $g$ is the determinant of the metric tensor. Suppose the initial state of the system is $|\psi_0\rangle=|G\rangle|G\rangle|0_M\rangle$, where $|G\rangle$ represents the ground state of each detector. The evolution operator is
\begin{equation}
  U=\mathcal{T}e^{-i\int d\tau H_I(\tau)},
\end{equation}
where $\mathcal{T}$ is the time-ordering operator. According to the perturbation theory, it can be written as
\begin{equation}
  U=1-i\int d\tau H_I(\tau)
  -\int d\tau_1\int^{\tau_1}d\tau_2H_I(\tau_1)H_I(\tau_2)+o(\lambda^2).
\end{equation}
The following steps are straigtfoward. We obtain the final state and subsequently the reduced density matrix of the detectors by tracing out the field state. Here we adopt the notations in ~\cite{Sachs-2017} and the form of the reduced density matrix is
\begin{equation}
  \begin{split}
    \rho=&\begin{pmatrix}
      1-P_A-P_B &   0   &   0   &   -M^*\\
      0         &   P_A &   L_{AB}^* & 0\\
      0         &   L_{AB} & P_B   & 0\\
      -M         &   0   &   0   &   0
    \end{pmatrix}+o(\lambda^2),
  \end{split}
\end{equation}
where
\begin{equation}
  P_j=\lambda_j^2\int d\tau_1\int d\tau_2  \chi_{j1} \chi_{j2}e^{i\Omega_j(\tau_2-\tau_1)}\int_{\Sigma_{j2}}\sqrt{-g_2}d^3\mathbf{x}_2f_{j2}\int_{\Sigma_{j1}}\sqrt{-g_1}d^3\mathbf{x}_1f_{j1}^*
    \langle0_M| {\mathcal{O}}_1 {\mathcal{O}}_2|0_M\rangle,
\end{equation}
with $j=A, B$, is just the transition probability for $A$ and $B$ detectors respectively,
\begin{align}
  L_{AB}=&\lambda_A\lambda_B\int d\tau_1\int d\tau_2  \chi_{A1} \chi_{B2} e^{-i\Omega_A\tau_1+i\Omega_B\tau_2}\int_{\Sigma_{A1}}\sqrt{-g_1}d^3\mathbf{x}_1f_{A1}^*\int_{\Sigma_{B2}}\sqrt{-g_2}d^3\mathbf{x}_2f_{B2} \langle0_M| {\mathcal{O}}_1 {\mathcal{O}}_2|0_M\rangle\label{eqn:LAB0}\\
  M=&\lambda_A\lambda_B\int d\tau_1\int d\tau_2\chi_{A1}\chi_{B2}e^{i\Omega_A\tau_1+i\Omega_B\tau_2}\int_{\Sigma_{A1}}f_{A1}\sqrt{-g_1}d^3\mathbf{x}_1\int_{\Sigma_{B2}}f_{B2}\sqrt{-g_2}d^3\mathbf{x}_2 \langle 0_M|\mathcal{T}\left( {\mathcal{O}}_1 {\mathcal{O}}_2\right)|0_M\rangle.\label{eqn:M0}
\end{align}

Our task now is to calculate these matrix  elements. We shall start with linear coupling as an illustration of the method,  then we  switch to quadratic coupling. Our method is to express the vacuum in terms of the states in the Fock space of Rindler modes. This means that we work in the reference frame of the accelerating observers. We will see that with this method,  divergence of the elements appears only when these two detectors concide in spacetime.

\section{Linear coupling}\label{sec:one_detector}

We start with the generic case and then focus on the special scenario in which the two detectors are identical. The main purpose  is to calculate the Wightman function.  Normally  one expands the field operator in Minkowski spacetime and integrates over the momentum variables to  obtain the Wightman function as a function of the spacetime coordinates. Then one could substitute the detectors' trajectories into the coordinates, over which one integrates in the end to obtain the matrix elements.

Here we use a different method. By  using the field expansion  \eqref{eqn: modeexpansion} and the expression of the Minkowski vaccum \eqref{eqn: thermofield-double}  in terms of Rindler modes,  we obtain
\begin{equation}
  \begin{split}
    &\langle 0_M| {\Phi}(\tau_1,\mathbf{x}_1) {\Phi}(\tau_2,\mathbf{x}_2)|0_M\rangle\\
    =&\prod_{\omega'\mathbf{k}'_\bot}\left(C_{\omega'}\sum_{n_{\omega'\mathbf{k}'_\bot}=0}^\infty e^{-\pi n_{\omega'\mathbf{k}'_\bot}\omega/a}\langle n_{\omega'\mathbf{k}'_\bot},R|\otimes\langle n_{\omega'\mathbf{k}'_\bot},L|\right)
    \int_0^\infty d\omega_1\iint d^2\mathbf{k}_{1\bot} \left( {a}^R_{\omega_1\mathbf{k}_{1\bot}}v^R_{1\omega_1\mathbf{k}_{1\bot}}+ {a}^{R\dag}_{\omega_1\mathbf{k}_{1\bot}}v^{R*}_{1\omega_1\mathbf{k}_{1\bot}}\right)\\
    &\times\int_0^\infty d\omega_2\iint d^2\mathbf{k}_{2\bot} \left( {a}^R_{\omega_2\mathbf{k}_{2\bot}}v^R_{2\omega_2\mathbf{k}_{2\bot}}+ {a}^{R\dag}_{\omega_2\mathbf{k}_{2\bot}}v^{R*}_{2\omega_2\mathbf{k}_{2\bot}}\right)
    \prod_{\omega\mathbf{k}_\bot}\left(C_\omega\sum_{n_{\omega\mathbf{k}_\bot}=0}^\infty e^{-\pi n_{\omega\mathbf{k}_\bot}\omega/a}|n_{\omega\mathbf{k}_\bot},R\rangle\otimes|n_{\omega\mathbf{k}_\bot},L\rangle\right)\\
    =&\int_0^\infty d\omega\iint d^2\mathbf{k}_{\bot} \left(v^R_{1\omega\mathbf{k}_{\bot}}v^{R*}_{2\omega\mathbf{k}_{\bot}}C_\omega^2\sum_{n=0}^\infty(n+1)e^{-2\pi n\omega/a}
    +v^{R*}_{1\omega\mathbf{k}_{\bot}}v^R_{2\omega\mathbf{k}_{\bot}}C_\omega^2  \sum_{n=1}^\infty ne^{-2\pi n\omega/a}\right)\\
    =&\int_0^\infty d\omega\iint d^2\mathbf{k}_{\bot} \left(v^R_{1\omega\mathbf{k}_{\bot}}v^{R*}_{2\omega\mathbf{k}_{\bot}}
    +v^{R*}_{1\omega\mathbf{k}_{\bot}}v^R_{2\omega\mathbf{k}_{\bot}}
    e^{-2\pi\omega/a}\right)\frac{1}{1-e^{-2\pi\omega/a}}.
    \label{eqn:WightmanFunction}
  \end{split}
\end{equation}
Note that our detectors reside in the same Rindler wedge. This result can also be seen directly from the expectation value of the Rindler number operator on the Minkowski vacuum.

Subsequently  we  do not integrate in the momentum space to obtain an explicit form of the Wightman function. It should also be pointed out that expanding the field with Rindler modes gives the same Wightman function as Minkowski modes~\cite{Crispino-2008}. Instead,  To calculate the matrix elements we  integrate over the spacetime coordinates, and thus express the elements as integrals over momentum variables. In some sense this method is more natural since we write quantities like transition probabilities as the sum of all different modes. The details are as follows.

\subsection{Calculation of $P$}

We can substitute the above equation into the transition probability and integrate over the spacetime variables, obtaining first
\begin{equation}
  \begin{split}
    P=&\lambda^2\int d\tau_1\int d\tau_2\chi(\tau_1)\chi(\tau_2)e^{-i\Omega(\tau_1-\tau_2)}\int_{\Sigma_2} f(\mathbf{x}_2)\sqrt{-g_2}d^3\mathbf{x}_2\int_{\Sigma_1} f(\mathbf{x}_1)^*\sqrt{-g_1}d^3\mathbf{x}_1\\
    &\times\int_0^\infty d\omega\iint d^2\mathbf{k}_{\bot} \left(v^R_{1\omega\mathbf{k}_{\bot}}v^{R*}_{2\omega\mathbf{k}_{\bot}}
    +v^{R*}_{1\omega\mathbf{k}_{\bot}}v^R_{2\omega\mathbf{k}_{\bot}}e^{-2\pi\omega/a}\right)\frac{1}{1-e^{-2\pi\omega/a}}\\
    =&\lambda^2\int_0^\infty d\omega\iint d^2\mathbf{k}_{\bot}\left(|\mu'_{\omega\mathbf{k}_{\bot}}|^2+|\mu_{\omega\mathbf{k}_{\bot}}|^2e^{-2\pi\omega/a}\right)\frac{1}{1-e^{-2\pi\omega/a}},\label{eqn:1_trans_prob_2}
  \end{split}
\end{equation}
where
\begin{equation}
  \begin{split}
    \mu_{\omega\mathbf{k}_{\bot}}=&\int \sqrt{-g}d\tau d\xi d^2\mathbf{x}_\bot\chi(\tau)v^R_{\omega\mathbf{k}_{\bot}}f(\xi,\mathbf{x}_\bot)e^{i\Omega\tau}\\
    =&\int d\tau d\xi d^2\mathbf{x}_\bot e^{2a\xi}\frac{1}{\sqrt{2\pi}}\int d\nu G(\nu)e^{-i\nu\tau}f(\xi,\mathbf{x}_\bot) \left[\frac{\sinh(\pi\omega/a)}{4\pi^4a}\right]^{1/2}K_{i\omega/a}\left(\frac{\kappa}{a}e^{a\xi}\right)e^{i\mathbf{k}_\bot\cdot\mathbf{x}_\bot-i\omega\tau+i\Omega\tau}\\
    =&\sqrt{2\pi}\int d\xi d^2\mathbf{x}_\bot e^{2a\xi}G(\Omega-\omega)f(\xi,\mathbf{x}_\bot) \left[\frac{\sinh(\pi\omega/a)}{4\pi^4a}\right]^{1/2}K_{i\omega/a}\left(\frac{\kappa}{a}e^{a\xi}\right)e^{i\mathbf{k}_\bot\cdot\mathbf{x}_\bot},\\
  \end{split}
\end{equation}
\begin{equation}
  \begin{split}
    \mu'_{\omega\mathbf{k}_{\bot}}=&\int \sqrt{-g}d\tau d\xi d^2\mathbf{x}_\bot\chi(\tau)v^{R*}_{\omega\mathbf{k}_{\bot}}f(\xi,\mathbf{x}_\bot)e^{i\Omega\tau}\\
    =&\int d\tau d\xi d^2\mathbf{x}_\bot e^{2a\xi}\chi(\tau)f(\xi,\mathbf{x}_\bot) \left[\frac{\sinh(\pi\omega/a)}{4\pi^4a}\right]^{1/2}K_{i\omega/a}\left(\frac{\kappa}{a}e^{a\xi}\right)e^{-i\mathbf{k}_\bot\cdot\mathbf{x}_\bot+i\omega\tau+i\Omega\tau}\\
    =&\int d\tau d\xi d^2\mathbf{x}_\bot e^{2a\xi}\frac{1}{\sqrt{2\pi}}\int d\nu G(\nu)e^{-i\nu\tau}f(\xi,\mathbf{x}_\bot) \left[\frac{\sinh(\pi\omega/a)}{4\pi^4a}\right]^{1/2}K_{i\omega/a}\left(\frac{\kappa}{a}e^{a\xi}\right)e^{-i\mathbf{k}_\bot\cdot\mathbf{x}_\bot+i\omega\tau+i\Omega\tau}\\
    =&\sqrt{2\pi}\int d\xi d^2\mathbf{x}_\bot e^{2a\xi}G(\Omega+\omega)f(\xi,\mathbf{x}_\bot) \left[\frac{\sinh(\pi\omega/a)}{4\pi^4a}\right]^{1/2}K_{i\omega/a}\left(\frac{\kappa}{a}e^{a\xi}\right)e^{-i\mathbf{k}_\bot\cdot\mathbf{x}_\bot}.\\
  \end{split}
\end{equation}
where $K$ is real,  $G(\nu)\equiv\frac{1}{\sqrt{2\pi}}\int d\tau\chi(\tau)e^{i\nu\tau}$ is the Fourier transformation of the switch function.

We are specifically interested in the case where the scalar field is massless and the detectors are point-like ones with Gaussian switch functions, which means
\begin{equation}
  \chi(\tau)=\frac{1}{\sqrt{2\pi}\sigma}
  \exp\left(-\frac{\tau^2}{2\sigma^2}\right),
  \,\,\, f(\xi,x,y)=\delta(\xi)\delta(x)\delta(y).
\end{equation}
Then the Fourier transformation takes the form
\begin{equation}
  G(\nu)=\exp\left(-\frac{\sigma^2}{2}\nu^2\right).
\end{equation}
Consequently,
\begin{align}
  &\mu_{\omega\mathbf{k}_{\bot}}=\sqrt{2\pi} \exp\left[-\frac{\sigma^2}{2}(\Omega-\omega)^2\right] \left[\frac{\sinh(\pi\omega/a)}{4\pi^4a}\right]^{1/2}K_{i\omega/a}\left(\frac{\kappa}{a}\right),\\
  &\mu'_{\omega\mathbf{k}_{\bot}}=\sqrt{2\pi} \exp\left[-\frac{\sigma^2}{2}(\Omega+\omega)^2\right] \left[\frac{\sinh(\pi\omega/a)}{4\pi^4a}\right]^{1/2}K_{i\omega/a}\left(\frac{\kappa}{a}\right),
\end{align}
Substitute the above equations into \eqref{eqn:1_trans_prob_2}, we obtain
\begin{equation}
  P=\frac{\lambda^2}{4\pi}\int_0^\infty d\omega \omega
  \frac{e^{-\sigma^2(\Omega+\omega)^2+\pi\omega/a}+e^{-\sigma^2(\Omega-\omega)^2-\pi\omega/a}}{\sinh(\pi\omega/a)}.
\end{equation}
The details are given in the appendix.

\subsection{Calculation of $L_{AB}$}

This is quite similar to the calculation of $P$, as can be easily seen from their expressions. We substitute the correlation function \eqref{eqn:WightmanFunction} into the expression   \eqref{eqn:LAB0} of   $L_{AB}$, obtaining
\begin{equation}
  \begin{split}
    L_{AB}=&\lambda_A\lambda_B\int d\tau_1\int d\tau_2  \chi_A(\tau_1) \chi_B(\tau_2) e^{-i\Omega_A\tau_1+i\Omega_B\tau_2}\int_{\Sigma_{A1}}\sqrt{-g_1}d^3\mathbf{x}_1f_A(\mathbf{x}_1)^*\int_{\Sigma_{B2}}\sqrt{-g_2}d^3\mathbf{x}_2f_B(\mathbf{x}_2)\\
    &\times\int_0^\infty d\omega\iint d^2\mathbf{k}_{\bot} \left(v^R_{1\omega\mathbf{k}_{\bot}}v^{R*}_{2\omega\mathbf{k}_{\bot}}
    +v^{R*}_{1\omega\mathbf{k}_{\bot}}v^R_{2\omega\mathbf{k}_{\bot}}e^{-2\pi\omega/a}\right)\frac{1}{1-e^{-2\pi\omega/a}}\\
    =&\lambda_A\lambda_B\int_0^\infty d\omega\iint d^2\mathbf{k}_{\bot}\left(\mu'^{A*}_{\omega\mathbf{k}_\bot}\mu'^{B}_{\omega\mathbf{k}_\bot}+\mu^{A*}_{\omega\mathbf{k}_\bot}\mu^{B}_{\omega\mathbf{k}_\bot}e^{-2\pi\omega/a}\right)\frac{1}{1-e^{-2\pi\omega/a}},  \label{eqn: lwithv}
  \end{split}
\end{equation}
where
\begin{equation}
  \begin{split}
    \mu^{j}_{\omega\mathbf{k}_\bot}=&\int \sqrt{-g}d\tau d\xi d^2\mathbf{x}_\bot\chi_j(\tau)v^R_{\omega\mathbf{k}_{\bot}}f_j(\xi,\mathbf{x}_\bot)e^{i\Omega_j\tau}\\
    =&\sqrt{2\pi}\int d\xi d^2\mathbf{x}_\bot e^{2a\xi}G_j(\Omega_j-\omega)f_j(\xi,\mathbf{x}_\bot) \left[\frac{\sinh(\pi\omega/a)}{4\pi^4a}\right]^{1/2}K_{i\omega/a}\left(\frac{\kappa}{a}e^{a\xi}\right)e^{i\mathbf{k}_\bot\cdot\mathbf{x}_\bot},
  \end{split}
\end{equation}
and
\begin{equation}
  \begin{split}
    \mu'^j_{\omega\mathbf{k}_{\bot}}=&\int \sqrt{-g}d\tau d\xi d^2\mathbf{x}_\bot\chi_j(\tau)v^{R*}_{\omega\mathbf{k}_{\bot}}f_j(\xi,\mathbf{x}_\bot)e^{i\Omega_j\tau}\\
    =&\sqrt{2\pi}\int d\xi d^2\mathbf{x}_\bot e^{2a\xi}G_j(\Omega_j+\omega)f_j(\xi,\mathbf{x}_\bot) \left[\frac{\sinh(\pi\omega/a)}{4\pi^4a}\right]^{1/2}K_{i\omega/a}\left(\frac{\kappa}{a}e^{a\xi}\right)e^{-i\mathbf{k}_\bot\cdot\mathbf{x}_\bot}.\\
  \end{split}
\end{equation}

Now we assume that these two detectors seperate in $x$ diretion with $x_0$ and there is a time delay $\tau_0$ between their switching functions, that is,
\begin{align}
  &\chi_A(\tau)=\frac{1}{\sqrt{2\pi}\sigma_A}\exp\left(-\frac{\tau^2}{2\sigma_A^2}\right),f_A(\xi,x,y)=\delta(\xi)\delta(x)\delta(y),\\
  &\chi_B(\tau)=\frac{1}{\sqrt{2\pi}\sigma_B}\exp\left[-\frac{(\tau-\tau_0)^2}{2\sigma_B^2}\right],f_B(\xi,x,y)=\delta(\xi)\delta(x-x_0)\delta(y).
\end{align}
Then we have (details are given in the appendix)
\begin{equation}
  \begin{split}
 L_{AB}=&\frac{\lambda_A\lambda_B}{\pi}\frac{1}{x_0\sqrt{a^2x_0^2+4}}
  \int_0^\infty d\omega\sin\frac{2\omega\text{arcsinh}(ax_0/2)}{a} e^{-\sigma_A^2(\Omega_A^2+\omega^2)/2-\sigma_B^2((\Omega_B^2+\omega^2)/2+i\Omega_B\tau_0}\\ &\times\frac{\cosh\left(\sigma_A^2\Omega_A\omega+\sigma_B^2\Omega_B\omega-i\omega\tau_0-\pi\omega/a\right)}{\sinh(\pi\omega/a)}.
 \end{split}
\end{equation}

\subsection{Calculation of $M$}

This is where things can get quite messy. We first calculate the time-ordered Minkowski vacuum expectation value
\begin{equation}
  \begin{split}
    &\langle 0_M|\mathcal{T}\left( {\Phi}_1 {\Phi}_2\right)|0_M\rangle\\
    =&\theta(\tau_1-\tau_2)\langle 0_M| {\Phi}(\tau_1,\mathbf{x}_1) {\Phi}(\tau_2,\mathbf{x}_2)|0_M\rangle +\theta(\tau_2-\tau_1)\langle 0_M| {\Phi}(\tau_2,\mathbf{x}_2) {\Phi}(\tau_1,\mathbf{x}_1)|0_M\rangle\\
    =&\theta(\tau_1-\tau_2)\int_0^\infty d\omega\iint d^2\mathbf{k}_{\bot} \left(v^R_{1\omega\mathbf{k}_{\bot}}v^{R*}_{2\omega\mathbf{k}_{\bot}}
    +v^{R*}_{1\omega\mathbf{k}_{\bot}}v^R_{2\omega\mathbf{k}_{\bot}}e^{-2\pi\omega/a}\right)\frac{1}{1-e^{-2\pi\omega/a}}\\
    &+\theta(\tau_2-\tau_1)\int_0^\infty d\omega\iint d^2\mathbf{k}_{\bot} \left(v^R_{2\omega\mathbf{k}_{\bot}}v^{R*}_{1\omega\mathbf{k}_{\bot}}
    +v^{R*}_{2\omega\mathbf{k}_{\bot}}v^R_{1\omega\mathbf{k}_{\bot}}e^{-2\pi\omega/a}\right)\frac{1}{1-e^{-2\pi\omega/a}},
  \end{split}
\end{equation}
in which $\theta(\tau)$ is the standard Heaviside function.
Substituting the above equation into \eqref{eqn:M0},  we obtain \begin{equation}
  \begin{split}
    M=&\lambda_A\lambda_B\int d\tau_1\int d\tau_2\chi_A(\tau_1)\chi_B(\tau_2)e^{i\Omega_A\tau_1+i\Omega_B\tau_2}\int_{\Sigma_{A1}}f_A(\mathbf{x}_1)\sqrt{-g_1}d^3\mathbf{x}_1\int_{\Sigma_{B2}}f_B(\mathbf{x}_2)\sqrt{-g_2}d^3\mathbf{x}_2\\
    &\times\left[\theta(\tau_1-\tau_2)\int_0^\infty d\omega\iint d^2\mathbf{k}_{\bot} \left(v^R_{1\omega\mathbf{k}_{\bot}}v^{R*}_{2\omega\mathbf{k}_{\bot}}
    +v^{R*}_{1\omega\mathbf{k}_{\bot}}v^R_{2\omega\mathbf{k}_{\bot}}e^{-2\pi\omega/a}\right)\frac{1}{1-e^{-2\pi\omega/a}}\right.\\
    &\left.+\theta(\tau_2-\tau_1)\int_0^\infty d\omega\iint d^2\mathbf{k}_{\bot} \left(v^R_{2\omega\mathbf{k}_{\bot}}v^{R*}_{1\omega\mathbf{k}_{\bot}}
    +v^{R*}_{2\omega\mathbf{k}_{\bot}}v^R_{1\omega\mathbf{k}_{\bot}}e^{-2\pi\omega/a}\right)\frac{1}{1-e^{-2\pi\omega/a}}\right]\\
    =&\lambda_A\lambda_B\int_0^\infty d\omega\iint d^2\mathbf{k}_{\bot}\left(W'_{\omega\mathbf{k}_{\bot}}
    +W_{\omega\mathbf{k}_{\bot}}e^{-2\pi\omega/a}\right)
    \frac{1}{1-e^{-2\pi\omega/a}},
  \end{split}
\end{equation}
with
\begin{equation}
  \begin{split}
    W_{\omega\mathbf{k}_{\bot}}=&\int d\tau_1\int d\tau_2\chi_A(\tau_1)\chi_B(\tau_2)e^{i\Omega_A\tau_1+i\Omega_B\tau_2}\int_{\Sigma_{A1}}f_A(\mathbf{x}_1)\sqrt{-g_1}d^3\mathbf{x}_1\int_{\Sigma_{B2}}f_B(\mathbf{x}_2)\sqrt{-g_2}d^3\mathbf{x}_2\\
    &\times\left[\theta(\tau_1-\tau_2)v^{R*}_{1\omega\mathbf{k}_{\bot}}v^R_{2\omega\mathbf{k}_{\bot}}+\theta(\tau_2-\tau_1)v^{R*}_{2\omega\mathbf{k}_{\bot}}v^R_{1\omega\mathbf{k}_{\bot}}\right]\\
    =&\int d\tau_1\int d\tau_2\chi_A(\tau_1)\chi_B(\tau_2)e^{i\Omega_A\tau_1+i\Omega_B\tau_2}\int_{\Sigma_{A1}}f_A(\mathbf{x}_1)\sqrt{-g_1}d^3\mathbf{x}_1\int_{\Sigma_{B2}}f_B(\mathbf{x}_2)\sqrt{-g_2}d^3\mathbf{x}_2\\
    &\frac{\sinh(\pi\omega/a)}{4\pi^4a}K_{i\omega/a}\left(\frac{\kappa}{a}e^{a\xi_1}\right)K_{i\omega/a}\left(\frac{\kappa}{a}e^{a\xi_2}\right) \frac{2\omega}{2\pi i}\int_{-\infty}^\infty du\frac{e^{iu(\tau_1-\tau_2)-i\mathbf{k}_\bot\cdot(\mathbf{x}_{1\bot}-\mathbf{x}_{2\bot})}}{u^2-\omega^2-i\epsilon}\\
    =&2\omega\frac{\sinh(\pi\omega/a)}{2\pi^3a(2\pi i)}\int du\frac{1}{u^2-\omega^2-i\epsilon}
    G_A(u+\Omega_A)\int d\xi_1d^2\mathbf{x}_{1\bot}e^{2a\xi_1}f_A(\xi_1,\mathbf{x}_{1\bot}) K_{i\omega/a}\left(\frac{\kappa}{a}e^{a\xi_1}\right)\\
    &\times G_B(\Omega_B-u)\int d\xi_2d^2\mathbf{x}_{2\bot}e^{2a\xi_2}f_B(\xi_2,\mathbf{x}_{2\bot}) K_{i\omega/a}\left(\frac{\kappa}{a}e^{a\xi_2}\right) e^{-i\mathbf{k}_\bot\cdot(\mathbf{x}_{1\bot}-\mathbf{x}_{2\bot})},
  \end{split}
\end{equation}
where $\epsilon$ is a positive infinitesimal real number.
Similarly we can obtain
\begin{equation}
  \begin{split}
    W'_{\omega\mathbf{k}_{\bot}}=&-2\omega\frac{\sinh(\pi\omega/a)}{2\pi^3a(2\pi i)}\int du\frac{1}{u^2-\omega^2+i\epsilon} G_A(u+\Omega_A)\int d\xi_1d^2\mathbf{x}_{1\bot}e^{2a\xi_1}f_A(\xi_1,\mathbf{x}_{1\bot}) K_{i\omega/a}\left(\frac{\kappa}{a}e^{a\xi_1}\right)\\
    &\times G_B(\Omega_B-u)\int d\xi_2d^2\mathbf{x}_{2\bot}e^{2a\xi_2}f_B(\xi_2,\mathbf{x}_{2\bot}) K_{i\omega/a}\left(\frac{\kappa}{a}e^{a\xi_2}\right) e^{-i\mathbf{k}_\bot\cdot(\mathbf{x}_{1\bot}-\mathbf{x}_{2\bot})}.
  \end{split}
\end{equation}

To further simplify the expressions, we assume  $\Omega_A=\Omega_B=\Omega,\sigma_A=\sigma_B=\sigma$ and $\tau_0=0$. Then we obtain
\begin{align}
  \begin{split}
    W_{\omega\mathbf{k}_\bot}=&4\omega\frac{\sinh(\pi\omega/a)}{2\pi^3a(2\pi i)}e^{-\sigma^2\Omega^2}K_{i\omega/a}\left(\frac{\kappa}{a}\right)^2 e^{i k_x x_0} \int_0^{\infty} du\frac{e^{-\sigma^2 u^2}}{u^2-\omega^2-i\epsilon}, \label{eqn: w1beforeana}
  \end{split}\\
  \begin{split}
    W'_{\omega\mathbf{k}_\bot}=&-4\omega\frac{\sinh(\pi\omega/a)}{2\pi^3a(2\pi i)}e^{-\sigma^2\Omega^2}K_{i\omega/a}\left(\frac{\kappa}{a}\right)^2 e^{i k_x x_0} \int_0^{\infty} du\frac{e^{-\sigma^2 u^2}}{u^2-\omega^2+i\epsilon}, \label{eqn: w2beforeana}
  \end{split}
\end{align}
The details are given in the Appendix. The final results are
\begin{align}
  \begin{split}
    W_{\omega\mathbf{k}_\bot}=&-\frac{\sinh(\pi\omega/a)}{2\pi^3a}e^{-\sigma^2\Omega^2}K_{i\omega/a}\left(\frac{\kappa}{a}\right)^2 e^{i k_x x_0} e^{-\sigma^2\omega^2}\text{erfc}(i\sigma\omega-2),
  \end{split}\\
  \begin{split}
    W'_{\omega\mathbf{k}_\bot}=&\frac{\sinh(\pi\omega/a)}{2\pi^3a}e^{-\sigma^2\Omega^2}K_{i\omega/a}\left(\frac{\kappa}{a}\right)^2 e^{i k_x x_0} e^{-\sigma^2\omega^2}\text{erfc}(i\sigma\omega).
  \end{split}
\end{align}
Substitute these into $M$, and we obtain
\begin{equation}
  \begin{split}
    M=&\lambda_A\lambda_B\int_0^\infty d\omega\iint d^2\mathbf{k}_{\bot}\left(W'_{\omega\mathbf{k}_{\bot}}+W_{\omega\mathbf{k}_{\bot}}e^{-2\pi\omega/a}\right)\frac{1}{1-e^{-2\pi\omega/a}}\\
    =&\frac{\lambda_A\lambda_Be^{-\sigma^2\Omega^2}}{2\pi^3a}\int_0^\infty d\omega\sinh(\pi\omega/a)e^{-\sigma^2\omega^2}\text{erfc}(i\sigma\omega) \iint d^2\mathbf{k}_{\bot}K_{i\omega/a}\left(\frac{\kappa}{a}\right)^2e^{i k_x x_0}\\
    =&\frac{\lambda_A\lambda_Be^{-\sigma^2\Omega^2}}{2\pi^3a}\int_0^\infty d\omega\sinh(\pi\omega/a)e^{-\sigma^2\omega^2}\text{erfc}(i\sigma\omega)\int_0^\infty k_\bot dk_\bot K_{i\omega/a}\left(\frac{\sqrt{k_\bot^2+m^2}}{a}\right)^2\int_0^{2\pi} d\alpha e^{ik_\bot x_0\cos\alpha}\\
    =&\frac{\lambda_A\lambda_Be^{-\sigma^2\Omega^2}}{\pi^2a}\int_0^\infty d\omega\sinh(\pi\omega/a)e^{-\sigma^2\omega^2}\text{erfc}(i\sigma\omega)\int_0^\infty k_\bot dk_\bot K_{i\omega/a}\left(\frac{\sqrt{k_\bot^2+m^2}}{a}\right)^2 J_0(k_\bot x_0)\\
    =&\frac{\lambda_A\lambda_Be^{-\sigma^2\Omega^2}}{\pi x_0\sqrt{a^2x_0^2+4}}\int_0^\infty d\omega e^{-\sigma^2\omega^2}\bigg[\text{erfc}(i\sigma\omega)+\frac{2e^{-2\pi\omega/a}}{1-e^{-2\pi\omega/a}}\bigg]\sin\frac{2\omega\text{arcsinh}(ax_0/2)}{a}.
  \end{split}
\end{equation}
where in the equality we've taken the field to be massless. 

Here is where the divergence may occur. If and only if two detectors concide in spacetime, i.e. $x_0=0$, $M$ is divergent:
\begin{equation}
  \begin{split}
    M=&\frac{\lambda^2}{2\pi}e^{-\sigma^2\Omega^2}\int_0^\infty d\omega \omega e^{-\sigma^2\omega^2}\bigg[\text{erfc}(i\sigma\omega)+\frac{2e^{-2\pi\omega/a}}{1-e^{-2\pi\omega/a}}\bigg].
  \end{split}
\end{equation}

\subsection{Calculation of Negativity}

\begin{figure}[ht!]
  \centering
  \includegraphics[scale=0.6]{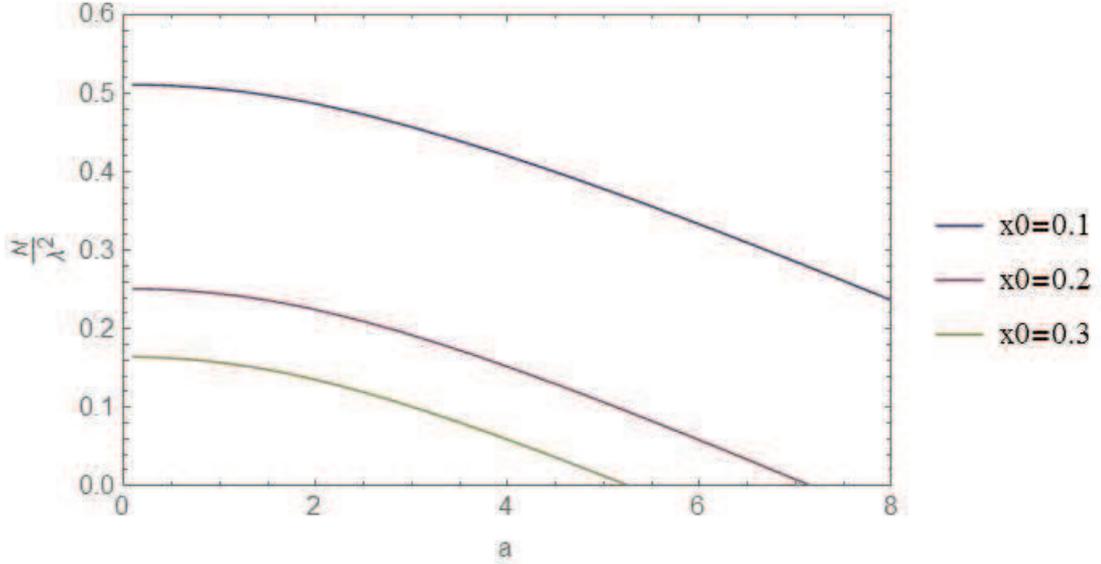}\\
  \caption{The graph for the negativity of the two detectors linearly coupled with the scalar field versus the acceleration ($\sigma=1, \Omega=1$)}\label{fig:NegativityLinear}
  \label{figneg}
\end{figure}

Now we can calculate the entanglement between the two detectors, with  $m=0,\lambda_A=\lambda_B=\lambda,\Omega_A=\Omega_B=\Omega,\sigma_A=\sigma_B=\sigma,\tau_0=0,x_0\neq0$. The matrix elements are obtained as
\begin{align}
    M=&\frac{\lambda^2}{\pi x_0\sqrt{a^2x_0^2+4}}e^{-\sigma^2\Omega^2}\int_0^\infty d\omega e^{-\sigma^2\omega^2}\bigg[\text{erfc}(i\sigma\omega)+\frac{2e^{-2\pi\omega/a}}{1-e^{-2\pi\omega/a}}\bigg]\sin\frac{2\omega\text{arcsinh}(ax_0/2)}{a},\\
    L_{AB}=&\frac{\lambda^2}{\pi x_0\sqrt{a^2x_0^2+4}}e^{-\sigma^2\Omega^2}
  \int_0^\infty d\omega e^{-\sigma^2\omega^2}\sin\frac{2\omega\text{arcsinh}(ax_0/2)}{a}\frac{\cosh\left(2\sigma^2\Omega\omega-\pi\omega/a\right)}{\sinh(\pi\omega/a)},\\
    P=&P_A=P_B=\frac{\lambda^2}{2\pi}e^{-\sigma^2\Omega^2}\int_0^\infty d\omega \omega e^{-\sigma^2\omega^2}\frac{\cosh(2\sigma^2\Omega\omega-\pi\omega/a)}{\sinh(\pi\omega/a)}.
\end{align}
The reduced density matrix has already been given in Sec.~2,
\begin{equation}
  \rho=\begin{pmatrix}
      1-2P &   0   &   0   &   -M^*\\
      0         &   P &   L_{AB} & 0\\
      0         &   L_{AB} & P   & 0\\
      -M         &   0   &   0   &   0
    \end{pmatrix}+o(\lambda^2).
\end{equation}
The partial transpose of it with respect to qubit A is
\begin{equation}
  \rho^{TA}=\begin{pmatrix}
      1-2P &   0   &   0   &   L_{AB}\\
      0         &   P &   -M^* & 0\\
      0         &   -M & P   & 0\\
       L_{AB}        &   0   &   0   &   0
    \end{pmatrix}+o(\lambda^2),
\end{equation}
with the four eigenvalues
\begin{align}
  \lambda_1=&\frac{1-2P+\sqrt{(1-2P)^2+4L^2}}{2}>0,\\
  \lambda_2=&\frac{1-2P-\sqrt{(1-2P)^2+4L^2}}{2}\propto\lambda^4,\\
  \lambda_3=&P+|M|>0,\\
  \lambda_4=&P-|M|.
\end{align}
Thus  the negativity of the reduced density matrix of eahc detector  is
\begin{equation}
  \mathcal{N}=\max(|M|-P,0).
\end{equation}
We obtained the numerical result  as   in Fig.~\ref{figneg}. 

It can be seen  that the negativity decreases with the acceleration and comes to zero at a certain point. In addition, the entanglement decreases with the increase of  the distance between the detectors, as anticipated.

\section{Quadratic coupling}

Now we consider two detectors which are quadratically coupled with a scalar field. The Hamiltonian in the interaction picture is

\begin{equation}
  H=H_A+H_B,
\end{equation}
where
\begin{equation}
  H_j=\lambda_j\chi_j(\tau)\int_{\Sigma_j}N\left[ {\Phi}(\tau,\mathbf{x})^2\right]\left[f_j(\mathbf{x})\sigma_j^+
  e^{i\Omega_j\tau}+f_j(\mathbf{x})^*\sigma_j^-
  e^{-i\Omega_j\tau}\right]\sqrt{-g}d^3\mathbf{x},
\end{equation}
j=A,B. Here  we take the field operator to be Wick ordered to avoid potential divergences. There are two equivalent ways to realize Wick ordering, one being putting the creation operators to the left of the annihilation operators, the other one being taking the field to be the  original one minus its vacuum expectation value. Both ways may cause ambiguity since the notion of mode expansion and the vacuum state is not uniquely defined in quantum field theory in curved spacetime. In this paper, we use the latter way.
In this way, the   Wightman function for the quadratic coupling is related to the that for the  linear coupling, which has been calculated in the last section.

The reduced density matrix shares the same structure with the case of $ {\Phi}$ coupling,
\begin{equation}
  \rho=\begin{pmatrix}
      1-P_A-P_B &   0   &   0   &   -M^*\\
      0         &   P_A &   L_{AB}^* & 0\\
      0         &   L_{AB} & P_B   & 0\\
      -M         &   0   &   0   &   0
    \end{pmatrix}+o(\lambda^2),
\end{equation}
where the matrix elements are
\begin{align}
  P_j=&\lambda_j^2\int d\tau_1\int d\tau_2  \chi_{j1} \chi_{j2}e^{i\Omega_j(\tau_2-\tau_1)}\int_{\Sigma_{j2}}\sqrt{-g_2}d^3\mathbf{x}_2f_{j2}\int_{\Sigma_{j1}}\sqrt{-g_1}d^3\mathbf{x}_1f_{j1}^*
    \langle0_M|N\left( {\Phi}_1^2\right)N\left( {\Phi}_2^2\right)|0_M\rangle,j=A,B, \\
  L_{AB}=&\lambda_A\lambda_B\int d\tau_1\int d\tau_2  \chi_{A1} \chi_{B2} e^{-i\Omega_A\tau_1+i\Omega_B\tau_2}\int_{\Sigma_{A1}}\sqrt{-g_1}d^3\mathbf{x}_1f_{A1}^*\int_{\Sigma_{B2}}\sqrt{-g_2}d^3\mathbf{x}_2f_{B2} \langle0_M|N\left( {\Phi}_1^2\right)N\left( {\Phi}_2^2\right)|0_M\rangle,\\
  M=&\lambda_A\lambda_B\int d\tau_1\int d\tau_2\chi_{A1}\chi_{B2}e^{i\Omega_A\tau_1+i\Omega_B\tau_2}\int_{\Sigma_{A1}}f_{A1}\sqrt{-g_1}d^3\mathbf{x}_1\int_{\Sigma_{B2}}f_{B2}\sqrt{-g_2}d^3\mathbf{x}_2 \langle 0_M|\mathcal{T}\left[N\left( {\Phi}_1^2\right)N\left( {\Phi}_2^2\right)\right]|0_M\rangle.
\end{align}
It should be clear now that we are to calculate the vacuum expectation value $V_{\Phi^2}$. It is closely related to the Wightman function, and we use \eqref{eqn:WightmanFunction} to obtain
\begin{equation}
  \begin{split}
    V_{\Phi^2}=&\langle0_M|N\left( {\Phi}_1^2\right)N\left( {\Phi}_2^2\right)|0_M\rangle\\
    =&2\langle0_M| {\Phi}_1 {\Phi}_2|0_M\rangle^2\\
    =&2\int_0^\infty d\omega_1\int_0^\infty d\omega_2\iint d^2\mathbf{k}_{1\bot}
    \iint d^2\mathbf{k}_{2\bot} \frac{1}{1-e^{-2\pi\omega_1/a}}\frac{1}{1-e^{-2\pi\omega_2/a}}\\
    &\times\left(v^R_{1\omega_1\mathbf{k}_{1\bot}}v^R_{1\omega_2\mathbf{k}_{2\bot}}v^{R*}_{2\omega_1\mathbf{k}_{1\bot}}v^{R*}_{2\omega_2\mathbf{k}_{2\bot}} +v^R_{1\omega_1\mathbf{k}_{1\bot}}v^{R*}_{1\omega_2\mathbf{k}_{2\bot}}v^{R*}_{2\omega_1\mathbf{k}_{1\bot}}v^R_{2\omega_2\mathbf{k}_{2\bot}}e^{-2\pi\omega_2/a} \right.\\ &\left.+v^{R*}_{1\omega_1\mathbf{k}_{1\bot}}v^R_{1\omega_2\mathbf{k}_{2\bot}}v^R_{2\omega_1\mathbf{k}_{1\bot}}v^{R*}_{2\omega_2\mathbf{k}_{2\bot} }e^{-2\pi\omega_1/a} +v^{R*}_{1\omega_1\mathbf{k}_{1\bot}}v^{R*}_{1\omega_2
    \mathbf{k}_{2\bot}}v^R_{2\omega_1\mathbf{k}_{1\bot}}
    v^R_{2\omega_2\mathbf{k}_{2\bot}}e^{-2\pi(\omega_1+
    \omega_2)/a}\right).
  \end{split}
\end{equation}

Now it is a good time to  make a digression to say a few words about the general structure of the interaction between detectors and quantum fields. The structure is well studied in ~\cite{Hummer-2016},   which  gives  the Feynman rules for the amplitude. In our case some clues of the Feynman rules can be observed. For example, for the quadratic coupling there are two vertices that give rise to  double integrals over the momentum variables. Besides, in our case, when the detectors start unentangled, the correlation term $M$ represents  a field propagator.   In genric cases (for instance the detectors star entangled), however, there are terms that represent detector propagators  . $\mathcal{T}\left[e^{-i\Omega (\tau_1-\tau_2)}\right]$. We wish to have a deeper insight to such terms in future works.

We follow the same procedure as in section 3 to calculate the matrix elements. The approach is the same but the calculation could be tedious so we leave the details in the Appendix and only  describe   the main steps and results.

\subsection{Calculation of $P_j$ and $L_{AB}$}

Let's start with $P_j$ and $L_{AB}$, since they are quite similar,
\begin{align}
    \begin{split}
  P_j=&2\lambda_j^2\int_0^\infty d\omega_1\int_0^\infty d\omega_2\iint d^2\mathbf{k}_{1\bot}
    \iint d^2\mathbf{k}_{2\bot}\frac{1}{1-e^{-2\pi\omega_1/a}}\frac{1}{1-e^{-2\pi\omega_2/a}}\\
    &\times\left(|\eta^{0j}_{\omega_1\mathbf{k}_{1\bot},\omega_2\mathbf{k}_{2\bot}}|^2 +|\eta^{2j}_{\omega_1\mathbf{k}_{1\bot},\omega_2\mathbf{k}_{2\bot}}|^2e^{-2\pi\omega_2/a} +|\eta^{1j}_{\omega_1\mathbf{k}_{1\bot},\omega_2\mathbf{k}_{2\bot}}|^2e^{-2\pi\omega_1/a}
    +|\eta^{12j}_{\omega_1\mathbf{k}_{1\bot},\omega_2\mathbf{k}_{2\bot}}|^2e^{-2\pi(\omega_1+\omega_2)/a}\right)\label{eqn:2_quar_Pj0},
 \end{split}\\
 \begin{split}
  L_{AB}=&2\lambda_A\lambda_B\int_0^\infty d\omega_1\int_0^\infty d\omega_2\iint d^2\mathbf{k}_{1\bot}
    \iint d^2\mathbf{k}_{2\bot} \frac{1}{1-e^{-2\pi\omega_1/a}}\frac{1}{1-e^{-2\pi\omega_2/a}}\\
    &\times\left(\eta^{0A*}_{\omega_1\mathbf{k}_{1\bot},\omega_2\mathbf{k}_{2\bot}}\eta^{0B}_{\omega_1\mathbf{k}_{1\bot},\omega_2\mathbf{k}_{2\bot}} +\eta^{2A*}_{\omega_1\mathbf{k}_{1\bot},\omega_2\mathbf{k}_{2\bot}}\eta^{2B}_{\omega_1\mathbf{k}_{1\bot},\omega_2\mathbf{k}_{2\bot}}e^{-2\pi\omega_2/a} +\eta^{1A*}_{\omega_1\mathbf{k}_{1\bot},\omega_2\mathbf{k}_{2\bot}}\eta^{1B}_{\omega_1\mathbf{k}_{1\bot},\omega_2\mathbf{k}_{2\bot}}e^{-2\pi\omega_1/a}\right.\\ &\left.+\eta^{12A*}_{\omega_1\mathbf{k}_{1\bot},\omega_2\mathbf{k}_{2\bot}}\eta^{12B}_{\omega_1\mathbf{k}_{1\bot},\omega_2\mathbf{k}_{2\bot}}e^{-2\pi(\omega_1+\omega_2)/a}\right)\label{eqn:2_quar_LAB0}.
   \end{split}
\end{align}

It is straightforward  to write down the form of the $\eta$'s. For instance,
\begin{equation}
  \begin{split}
    \eta^0_{\omega_1\mathbf{k}_{1\bot},\omega_2\mathbf{k}_{2\bot}}=&\int d\tau \chi(\tau)e^{i\Omega\tau}\int_\Sigma\sqrt{-g}d^3\mathbf{x}f(\mathbf{x}) v^{R*}_{\omega_1\mathbf{k}_{1\bot}}v^{R*}_{\omega_2\mathbf{k}_{2\bot}}\\
    =&\int d\tau d\xi d^2\mathbf{x}_\bot \chi(\tau) e^{2a\xi}f(\xi,\mathbf{x}_\bot) \left[\frac{\sinh(\pi\omega_1/a)}{4\pi^4a}\right]^{1/2}K_{i\omega_1/a}\left(\frac{\kappa_1}{a}e^{a\xi}\right)e^{-i\mathbf{k}_{1\bot}\cdot\mathbf{x}_\bot+i\omega_1\tau}\\ &\times\left[\frac{\sinh(\pi\omega_2/a)}{4\pi^4a}\right]^{1/2}K_{i\omega_2/a}\left(\frac{\kappa_2}{a}e^{a\xi}\right)e^{-i\mathbf{k}_{2\bot}\cdot\mathbf{x}_\bot+i\omega_2\tau}e^{i\Omega\tau}\\
    =&\int d\tau d\xi d^2\mathbf{x}_\bot \frac{1}{\sqrt{2\pi}}\int d\nu G(\nu)e^{-i\nu\tau} e^{2a\xi}f(\xi,\mathbf{x}_\bot) e^{-i(\mathbf{k}_{1\bot}+\mathbf{k}_{2\bot})\cdot\mathbf{x}_\bot+i(\omega_1+\omega_2+\Omega)\tau} \\
     &\times \left[\frac{\sinh(\pi\omega_1/a)}{4\pi^4a}\right]^{1/2}K_{i\omega_1/a}\left(\frac{\kappa_1}{a}e^{a\xi}\right) \left[\frac{\sinh(\pi\omega_2/a)}{4\pi^4a}\right]^{1/2}K_{i\omega_2/a}\left(\frac{\kappa_2}{a}e^{a\xi}\right)\\
    =&\frac{\sqrt{2\pi}}{4\pi^4a}\int d\xi d^2\mathbf{x}_\bot G(\omega_1+\omega_2+\Omega)e^{2a\xi}f(\xi,\mathbf{x}_\bot) e^{-i(\mathbf{k}_{1\bot}+\mathbf{k}_{2\bot})\cdot\mathbf{x}_\bot}\\
    &\times\sqrt{\sinh(\pi\omega_1/a)\sinh(\pi\omega_2/a)}K_{i\omega_1/a}\left(\frac{\kappa_1}{a}e^{a\xi}\right)K_{i\omega_2/a}\left(\frac{\kappa_2}{a}e^{a\xi}\right).
  \end{split}
\end{equation}

The final expression for  the parameter values $m=0,\lambda_A=\lambda_B=\lambda,
\Omega_A=\Omega_B=\Omega,\sigma_A=\sigma_B=\sigma,
\tau_0=0,x_0\neq0$ will be given later along with $M$.

\subsection{Calculation of $M$}

Now we calculate $M$, which was divergent in previous calculations by other other authors. We need to calculate the time-ordered vacuum expectation value
\begin{equation}
  \begin{split}
    D_{\Phi^2}=&\langle 0_M|\mathcal{T}\left[N\left( {\Phi}_1^2\right)N\left( {\Phi}_2^2\right)\right]|0_M\rangle\\
    =&2\theta(\tau_1-\tau_2)\langle 0_M| {\Phi}(\tau_1,\mathbf{x}_1) {\Phi}(\tau_2,\mathbf{x}_2)|0_M\rangle^2 +2\theta(\tau_2-\tau_1)\langle 0_M| {\Phi}(\tau_2,\mathbf{x}_2) {\Phi}(\tau_1,\mathbf{x}_1)|0_M\rangle^2\\
    =&2\int_0^\infty d\omega_1\int_0^\infty d\omega_2\iint d^2\mathbf{k}_{1\bot}
    \iint d^2\mathbf{k}_{2\bot} \frac{1}{1-e^{-2\pi\omega_1/a}}\frac{1}{1-e^{-2\pi\omega_2/a}}\\
    &\times\left[\theta(\tau_1-\tau_2)\left(v^R_{1\omega_1\mathbf{k}_{1\bot}}v^R_{1\omega_2\mathbf{k}_{2\bot}}v^{R*}_{2\omega_1\mathbf{k}_{1\bot}}v^{R*}_{2\omega_2\mathbf{k}_{2\bot}} +v^R_{1\omega_1\mathbf{k}_{1\bot}}v^{R*}_{1\omega_2\mathbf{k}_{2\bot}}v^{R*}_{2\omega_1\mathbf{k}_{1\bot}}v^R_{2\omega_2\mathbf{k}_{2\bot}}e^{-2\pi\omega_2/a} \right.\right.\\
    &\left.+v^{R*}_{1\omega_1\mathbf{k}_{1\bot}}v^R_{1\omega_2\mathbf{k}_{2\bot}}v^R_{2\omega_1\mathbf{k}_{1\bot}}v^{R*}_{2\omega_2\mathbf{k}_{2\bot} }e^{-2\pi\omega_1/a} +v^{R*}_{1\omega_1\mathbf{k}_{1\bot}}v^{R*}_{1\omega_2\mathbf{k}_{2\bot}}v^R_{2\omega_1\mathbf{k}_{1\bot}}v^R_{2\omega_2\mathbf{k}_{2\bot}}e^{-2\pi(\omega_1+\omega_2)/a}\right)\\
    &+\theta(\tau_2-\tau_1)\left(v^R_{2\omega_1\mathbf{k}_{1\bot}}v^R_{2\omega_2\mathbf{k}_{2\bot}}v^{R*}_{1\omega_1\mathbf{k}_{1\bot}}v^{R*}_{1\omega_2\mathbf{k}_{2\bot}} +v^R_{2\omega_1\mathbf{k}_{1\bot}}v^{R*}_{2\omega_2\mathbf{k}_{2\bot}}v^{R*}_{1\omega_1\mathbf{k}_{1\bot}}v^R_{1\omega_2\mathbf{k}_{2\bot}}e^{-2\pi\omega_2/a} \right.\\
    &\left.\left.+v^{R*}_{2\omega_1\mathbf{k}_{1\bot}}v^R_{2\omega_2\mathbf{k}_{2\bot}}v^R_{1\omega_1\mathbf{k}_{1\bot}}v^{R*}_{1\omega_2\mathbf{k}_{2\bot} }e^{-2\pi\omega_1/a} +v^{R*}_{2\omega_1\mathbf{k}_{1\bot}}v^{R*}_{2\omega_2\mathbf{k}_{2\bot}}v^R_{1\omega_1\mathbf{k}_{1\bot}}v^R_{1\omega_2\mathbf{k}_{2\bot}}e^{-2\pi(\omega_1+\omega_2)/a}\right)\right].
  \end{split}
\end{equation}

Substituting this expression into $M$, we obtain
\begin{equation}
  \begin{split}
    M=&2\lambda_A\lambda_B\int d\tau_1\int d\tau_2\chi_{A1}\chi_{B2}e^{i\Omega_A\tau_1+i\Omega_B\tau_2}\int_{\Sigma_{A1}}f_{A1}\sqrt{-g_1}d^3\mathbf{x}_1\int_{\Sigma_{B2}}f_{B2}\sqrt{-g_2}d^3\mathbf{x}_2 \\
    &\int_0^\infty d\omega_1\int_0^\infty d\omega_2\iint d^2\mathbf{k}_{1\bot}
    \iint d^2\mathbf{k}_{2\bot} \frac{1}{1-e^{-2\pi\omega_1/a}}\frac{1}{1-e^{-2\pi\omega_2/a}}\\
    &\times\left[\theta(\tau_1-\tau_2)\left(v^R_{1\omega_1\mathbf{k}_{1\bot}}v^R_{1\omega_2\mathbf{k}_{2\bot}}v^{R*}_{2\omega_1\mathbf{k}_{1\bot}}v^{R*}_{2\omega_2\mathbf{k}_{2\bot}} +v^R_{1\omega_1\mathbf{k}_{1\bot}}v^{R*}_{1\omega_2\mathbf{k}_{2\bot}}v^{R*}_{2\omega_1\mathbf{k}_{1\bot}}v^R_{2\omega_2\mathbf{k}_{2\bot}}e^{-2\pi\omega_2/a} \right.\right.\\
    &\left.+v^{R*}_{1\omega_1\mathbf{k}_{1\bot}}v^R_{1\omega_2\mathbf{k}_{2\bot}}v^R_{2\omega_1\mathbf{k}_{1\bot}}v^{R*}_{2\omega_2\mathbf{k}_{2\bot} }e^{-2\pi\omega_1/a} +v^{R*}_{1\omega_1\mathbf{k}_{1\bot}}v^{R*}_{1\omega_2\mathbf{k}_{2\bot}}v^R_{2\omega_1\mathbf{k}_{1\bot}}v^R_{2\omega_2\mathbf{k}_{2\bot}}e^{-2\pi(\omega_1+\omega_2)/a}\right)\\
    &+\theta(\tau_2-\tau_1)\left(v^R_{2\omega_1\mathbf{k}_{1\bot}}v^R_{2\omega_2\mathbf{k}_{2\bot}}v^{R*}_{1\omega_1\mathbf{k}_{1\bot}}v^{R*}_{1\omega_2\mathbf{k}_{2\bot}} +v^R_{2\omega_1\mathbf{k}_{1\bot}}v^{R*}_{2\omega_2\mathbf{k}_{2\bot}}v^{R*}_{1\omega_1\mathbf{k}_{1\bot}}v^R_{1\omega_2\mathbf{k}_{2\bot}}e^{-2\pi\omega_2/a} \right.\\
    &\left.\left.+v^{R*}_{2\omega_1\mathbf{k}_{1\bot}}v^R_{2\omega_2\mathbf{k}_{2\bot}}v^R_{1\omega_1\mathbf{k}_{1\bot}}v^{R*}_{1\omega_2\mathbf{k}_{2\bot} }e^{-2\pi\omega_1/a} +v^{R*}_{2\omega_1\mathbf{k}_{1\bot}}v^{R*}_{2\omega_2\mathbf{k}_{2\bot}}v^R_{1\omega_1\mathbf{k}_{1\bot}}v^R_{1\omega_2\mathbf{k}_{2\bot}}e^{-2\pi(\omega_1+\omega_2)/a}\right)\right]\\
    =&2\lambda_A\lambda_B\int_0^\infty d\omega_1\int_0^\infty d\omega_2\iint d^2\mathbf{k}_{1\bot}
    \iint d^2\mathbf{k}_{2\bot} \frac{1}{1-e^{-2\pi\omega_1/a}}\frac{1}{1-e^{-2\pi\omega_2/a}}\\
    &\times\left[W^0_{\omega_1\mathbf{k}_{1\bot},\omega_2\mathbf{k}_{2\bot}} +W^2_{\omega_1\mathbf{k}_{1\bot},\omega_2\mathbf{k}_{2\bot}}e^{-2\pi\omega_2/a} +W^1_{\omega_1\mathbf{k}_{1\bot},\omega_2\mathbf{k}_{2\bot}}e^{-2\pi\omega_1/a} +W^{12}_{\omega_1\mathbf{k}_{1\bot},\omega_2\mathbf{k}_{2\bot}}e^{-2\pi(\omega_1+\omega_2)/a}\right].\label{eqn:2_quar_M0}
  \end{split}
\end{equation}
These terms  looks complicated but can be  calculated straightforwardly. Here we take $W^0_{\omega_1\mathbf{k}_{1\bot},\omega_2\mathbf{k}_{2\bot}}$ as an example.

\begin{align*}
\hspace{-2cm}W^0_{\omega_1\mathbf{k}_{1\bot},\omega_2\mathbf{k}_{2\bot}}=&\int d\tau_1\int d\tau_2\chi_{A1}\chi_{B2}e^{i\Omega_A\tau_1+i\Omega_B\tau_2}\int_{\Sigma_{A1}}f_{A1}\sqrt{-g_1}d^3\mathbf{x}_1\int_{\Sigma_{B2}}f_{B2}\sqrt{-g_2}d^3\mathbf{x}_2\\
    &\times\left[\theta(\tau_1-\tau_2)v^R_{1\omega_1\mathbf{k}_{1\bot}}v^R_{1\omega_2\mathbf{k}_{2\bot}}v^{R*}_{2\omega_1\mathbf{k}_{1\bot}}v^{R*}_{2\omega_2\mathbf{k}_{2\bot}} +\theta(\tau_2-\tau_1)v^R_{2\omega_1\mathbf{k}_{1\bot}}v^R_{2\omega_2\mathbf{k}_{2\bot}}v^{R*}_{1\omega_1\mathbf{k}_{1\bot}}v^{R*}_{1\omega_2\mathbf{k}_{2\bot}}\right]\\
    =&\frac{\sinh(\pi\omega_1/a)\sinh(\pi\omega_2/a)}{(4\pi^4a)^2}\int d\tau_1\int d\tau_2\chi_{A1}\chi_{B2}e^{i\Omega_A\tau_1+i\Omega_B\tau_2}\int_{\Sigma_{A1}}f_{A1}\sqrt{-g_1}d^3\mathbf{x}_1\int_{\Sigma_{B2}}f_{B2}\sqrt{-g_2}d^3\mathbf{x}_2\\
    &\times K_{i\omega_1/a}\left(\frac{\kappa_1}{a}e^{a\xi_1}\right) K_{i\omega_2/a}\left(\frac{\kappa_2}{a}e^{a\xi_1}\right)K_{i\omega_1/a}\left(\frac{\kappa_1}{a}e^{a\xi_2}\right) K_{i\omega_2/a}\left(\frac{\kappa_2}{a}e^{a\xi_2}\right)\\
    &\times\left[\theta(\tau_1-\tau_2)e^{i(\mathbf{k}_{1\bot}+\mathbf{k}_{2\bot})\cdot(\mathbf{x}_{1\bot}-\mathbf{x}_{2\bot})-i(\omega_1+\omega_2)(\tau_1-\tau_2)} +\theta(\tau_2-\tau_1)e^{-i(\mathbf{k}_{1\bot}+\mathbf{k}_{2\bot})\cdot(\mathbf{x}_{1\bot}-\mathbf{x}_{2\bot})+i(\omega_1+\omega_2)(\tau_1-\tau_2)}\right]\\
    =&\frac{\sinh(\pi\omega_1/a)\sinh(\pi\omega_2/a)}{(4\pi^4a)^2}\int d\tau_1\int d\tau_2\chi_{A1}\chi_{B2}e^{i\Omega_A\tau_1+i\Omega_B\tau_2}\int_{\Sigma_{A1}}f_{A1}\sqrt{-g_1}d^3\mathbf{x}_1\int_{\Sigma_{B2}}f_{B2}\sqrt{-g_2}d^3\mathbf{x}_2\\
    &\times K_{i\omega_1/a}\left(\frac{\kappa_1}{a}e^{a\xi_1}\right) K_{i\omega_2/a}\left(\frac{\kappa_2}{a}e^{a\xi_1}\right)K_{i\omega_1/a}\left(\frac{\kappa_1}{a}e^{a\xi_2}\right) K_{i\omega_2/a}\left(\frac{\kappa_2}{a}e^{a\xi_2}\right)\\
    &\times\frac{-2(\omega_1+\omega_2)}{2\pi i}\int du\frac{e^{iu(\tau_1-\tau_2)-i(\mathbf{k}_{1\bot}+\mathbf{k}_{2\bot})\cdot(\mathbf{x}_{1\bot}-\mathbf{x}_{2\bot})}}{u^2-(\omega_1+\omega_2)^2+i\epsilon}\\
    =&\frac{\sinh(\pi\omega_1/a)\sinh(\pi\omega_2/a)}{(4\pi^4a)^2/(2\pi)}\frac{-2(\omega_1+\omega_2)}{2\pi i}\int du \frac{G_A(\Omega_A+u)G_B(\Omega_B-u)}{u^2-(\omega_1+\omega_2)^2+i\epsilon}
    \int_{\Sigma_{A1}}f_{A1}\sqrt{-g_1}d^3\mathbf{x}_1\int_{\Sigma_{B2}}f_{B2}\sqrt{-g_2}d^3\mathbf{x}_2\\
    &\times K_{i\omega_1/a}\left(\frac{\kappa_1}{a}e^{a\xi_1}\right) K_{i\omega_2/a}\left(\frac{\kappa_2}{a}e^{a\xi_1}\right)K_{i\omega_1/a}\left(\frac{\kappa_1}{a}e^{a\xi_2}\right) K_{i\omega_2/a}\left(\frac{\kappa_2}{a}e^{a\xi_2}\right)
    e^{-i(\mathbf{k}_{1\bot}+\mathbf{k}_{2\bot})\cdot(\mathbf{x}_{1\bot}-\mathbf{x}_{2\bot})}.\\
\end{align*}
As before, keep in mind that in the end we will integrate over $\mathbf{k}_{1\bot}$ and $\mathbf{k}_{2\bot}$.
The integral over $u$ can be calculated by using the same technique as in the linear case,
\begin{align}
  \begin{split}
    W^0_{\omega_1\mathbf{k}_{1\bot},\omega_2\mathbf{k}_{2\bot}}=&\frac{\sinh(\pi\omega_1/a)\sinh(\pi\omega_2/a)}{(4\pi^4a)^2/(2\pi)} K_{i\omega_1/a}\left(\frac{\kappa_1}{a}\right)^2K_{i\omega_2/a}\left(\frac{\kappa_2}{a}\right)^2e^{-\sigma^2\Omega^2}\\
  &\times e^{i(k_{1x}+k_{2x})x_0}e^{-\sigma^2(\omega_1+\omega_2)^2}\text{erfc}[i\sigma(\omega_1+\omega_2)].
  \end{split}
\end{align}
The other $W$'s can be treated the similar way. Substituting the $W$s into $M$ and taking the field to be massless, one obtains
\begin{equation}
  \begin{split}
    M=&\frac{\lambda_A\lambda_B}{\pi^3x_0^2(a^2x_0^2+4)}e^{-\sigma^2\Omega^2}\int_0^\infty d\omega_1\int_0^\infty d\omega_2 \Bigg\{\frac{\sin\frac{2\omega_1\text{arcsinh}(ax_0/2)}{a}}{2\sinh(\pi\omega_1/a)} \frac{\sin\frac{2\omega_2\text{arcsinh}(ax_0/2)}{a}}{\sinh(\pi\omega_2/a)}\\
    &\times\left[e^{-\sigma^2(\omega_1+\omega_2)^2}\text{erfc}[i\sigma(\omega_1+\omega_2)]\sinh[\pi(\omega_1+\omega_2)/a] +e^{-\sigma^2(\omega_1-\omega_2)^2}\text{erfc}[i\sigma(\omega_1-\omega_2)]\sinh[\pi(\omega_1-\omega_2)/a]\right]\\
    &+\frac{\sin\frac{2\omega_1\text{arcsinh}(ax_0/2)}{a}\sin\frac{2\omega_2\text{arcsinh}(ax_0/2)}{a}}{(1-e^{-2\pi \omega_1/a})(1-e^{-2\pi \omega_2/a})}\times\big[e^{-\sigma^2(\omega_1-\omega_2)^2-2\pi\omega_2/a}(\text{sgn}(\omega_1-\omega_2)-1)\\
    &+e^{-\sigma^2(\omega_1-\omega_2)^2-2\pi\omega_1/a}(\text{sgn}(\omega_1-\omega_2)+1)+2e^{-\sigma^2 (\omega_1+\omega_2)^2-2\pi (\omega_1+\omega_2)/a}\big]\Bigg\}.
  \end{split}
\end{equation}

As in the linear case, $M$ is divergent when $x_0=0$:
\begin{equation}
  \begin{split}
    M=&\frac{\lambda^2}{4\pi^3}e^{-\sigma^2\Omega^2}\int_0^\infty d\omega_1\int_0^\infty d\omega_2 \Bigg\{\frac{\omega_1}{2\sinh(\pi\omega_1/a)} \frac{\omega_2}{\sinh(\pi\omega_2/a)}\\
    &\times\left[e^{-\sigma^2(\omega_1+\omega_2)^2}\text{erfc}[i\sigma(\omega_1+\omega_2)]\sinh[\pi(\omega_1+\omega_2)/a] +e^{-\sigma^2(\omega_1-\omega_2)^2}\text{erfc}[i\sigma(\omega_1-\omega_2)]\sinh[\pi(\omega_1-\omega_2)/a]\right]\\
    &+\frac{\omega_1\omega_2}{(1-e^{-2\pi \omega_1/a})(1-e^{-2\pi \omega_2/a})}\times\big[e^{-\sigma^2(\omega_1-\omega_2)^2-2\pi\omega_2/a}(\text{sgn}(\omega_1-\omega_2)-1)\\
    &+e^{-\sigma^2(\omega_1-\omega_2)^2-2\pi\omega_1/a}(\text{sgn}(\omega_1-\omega_2)+1)+2e^{-\sigma^2 (\omega_1+\omega_2)^2-2\pi (\omega_1+\omega_2)/a}\big]\Bigg\}.
  \end{split}
\end{equation}

\paragraph{Calculation of Negativity}

Now under the same parameter values as in section 3, i.e. $m=0,\lambda_A=\lambda_B=\lambda,\Omega_A=\Omega_B=\Omega,\sigma_A=\sigma_B=\sigma,\tau_0=0,x_0\neq0$, the elements of the reduced density matrix are
\begin{align}
  \begin{split}
    M=&\frac{\lambda^2}{\pi^3x_0^2(a^2x_0^2+4)}e^{-\sigma^2\Omega^2}\int_0^\infty d\omega_1\int_0^\infty d\omega_2 \Bigg\{\frac{\sin\frac{2\omega_1\text{arcsinh}(ax_0/2)}{a}}{2\sinh(\pi\omega_1/a)} \frac{\sin\frac{2\omega_2\text{arcsinh}(ax_0/2)}{a}}{\sinh(\pi\omega_2/a)}\\
    &\times\left[e^{-\sigma^2(\omega_1+\omega_2)^2}\text{erfc}[i\sigma(\omega_1+\omega_2)]\sinh[\pi(\omega_1+\omega_2)/a] +e^{-\sigma^2(\omega_1-\omega_2)^2}\text{erfc}[i\sigma(\omega_1-\omega_2)]\sinh[\pi(\omega_1-\omega_2)/a]\right]\\
    &+\frac{\sin\frac{2\omega_1\text{arcsinh}(ax_0/2)}{a}\sin\frac{2\omega_2\text{arcsinh}(ax_0/2)}{a}}{(1-e^{-2\pi \omega_1/a})(1-e^{-2\pi \omega_2/a})}\times\big[e^{-\sigma^2(\omega_1-\omega_2)^2-2\pi\omega_2/a}(\text{sgn}(\omega_1-\omega_2)-1)\\
    &+e^{-\sigma^2(\omega_1-\omega_2)^2-2\pi\omega_1/a}(\text{sgn}(\omega_1-\omega_2)+1)+2e^{-\sigma^2 (\omega_1+\omega_2)^2-2\pi (\omega_1+\omega_2)/a}\big]\Bigg\},
  \end{split}\\
  \begin{split}
    P=&\frac{\lambda^2}{16\pi^3}\int_0^\infty d\omega_1\int_0^\infty d\omega_2\frac{\omega_1}{\sinh(\pi\omega_1/a)}\frac{\omega_2}{\sinh(\pi\omega_2/a)}\\
    &\times\left\{\exp\left[-\sigma^2(\omega_1+\omega_2+\Omega)^2+\pi(\omega_1+\omega_2)/a\right] +\exp\left[-\sigma^2(\Omega+\omega_1-\omega_2)^2+\pi(\omega_1-\omega_2)/a\right]\right.\\ &\left.+\exp\left[-\sigma^2(\Omega-\omega_1+\omega_2)^2-\pi(\omega_1-\omega_2)/a\right]
    +\exp\left[-\sigma^2(\Omega-\omega_1-\omega_2)^2-\pi(\omega_1+\omega_2)/a\right]\right\}.
  \end{split}
\end{align}

As before, the negativity for the reduced density matrix of the detectors is
\begin{equation}
  \mathcal{N}=\max(|M|-P,0)
\end{equation}
and the numerical result is depicted in Figure 4.1.

\begin{figure}[ht!]
  \centering
  \includegraphics[scale=0.6]{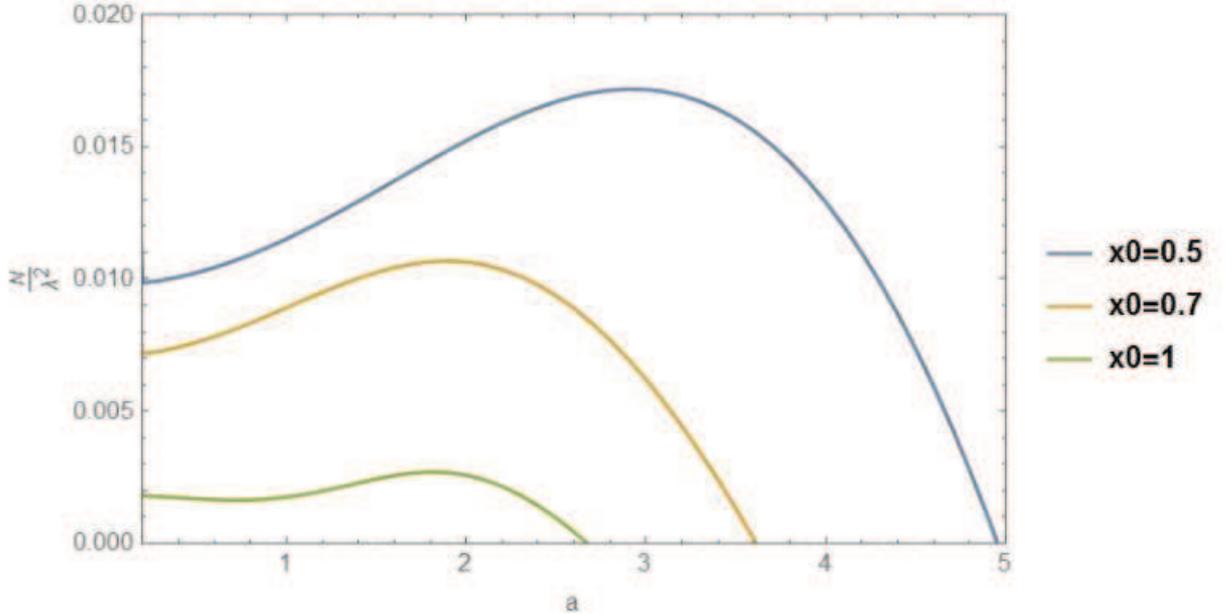}\\
  \caption{Negativity of the two detectors quadratically coupled with the scalar field, as a function of the acceleration ($\sigma=1, \tau=1$)}\label{fig:NegativityLinear}
\end{figure}

It can be seen that now the negativity can be enhanced by the acceleration, though it still vanishes when the acceleration is large enough. The entanglement also decreases with the increase of the  distance.

\section{Conclusion}

We have studied two Unruh-DeWitt detectors  coupled with a scalar field linearly or quadratically. Other than working in the inertial frame, as people usually do, we use  Rindler coordinates, as  the Hamiltonian itself depends on the proper time of the detectors. In our calculations,  the divergence people usually encounter in calculating  correlations  only appears  when the detectors coincide in spacetime. Moreover, the dependence of the entanglement on the acceleration is more complex in the quadratic-coupling case, and the dependence is not monotonic. In all cases studied,   the entanglement decreases with the increase of   the distance between the detectors, in  consistency  with our intuition.

To deal with the Wightman function, which is divergent   at certain points in spacetime, our approach is different from previous ones, and is based on the field expansion in the Rindler coordinates. Moreover,in calculating the  matrix elements of the reduced density matrix,   rather than making  integration  over the momentum variables immediately,  we first integrate over the spacetime coordinates. This may be the reason why we do not encounter divergence except when the detectors coincide in spacetime.

Given the equivalence between the Rindler spacetime and the Schwarzchild spacetime near the horizon, our results can be applied to the case of black hole. We hope that our method can extended to more    general cases. In our model, it is supposed that  the two detectors   reside in the same Rindler wedge to avoid ambiguities. Moreover, it is quite promising that we could apply our method to the case of Dirac interaction. Further extension can be made to consider the case that  the detectors are  entangled initially.

\appendix

\section{Special Integrals}

\begin{align}
  &\int_0^{2\pi}e^{\pm iu\cos\alpha}d\alpha=2\pi J_0(u)\\
  &\int_0^{2\pi}\cos\alpha e^{\pm iu\cos\alpha}d\alpha=\pm2\pi i J_1(u)\\
  &\int_0^\infty k K_{i\omega/a}(k/a)^2J_0(kx_0)dk=\frac{\pi a}{x_0\sqrt{a^2x_0^2+4}\sinh(\pi\omega/a)}\sin\frac{2\omega\text{arcsinh}(ax_0/2)}{a}\\
  &\int_0^\infty k^2 [K_{i\omega/a-1}(k/a)+K_{i\omega/a+1}(k/a)]K_{i\omega/a}(k/a)J_0(kx_0)dk=\\
  &\quad\frac{2a^2\pi}{(a^2x_0^2+4)^2\sinh(\pi\omega/a)}\left[(a^2x_0^2+4)\omega \cos\frac{2\omega\text{arcsinh}(ax_0/2)}{a}+\frac{2}{x_0}\sqrt{a^2x_0^2+4} \sin\frac{2\omega\text{arcsinh}(ax_0/2)}{a}\right]\\
  &\int_0^\infty k^2K_{i\omega/a}(k/a)^2J_1(kx_0)dk=\\
  &\frac{\frac{2\pi a}{\sinh(\pi\omega/a)}\left[-\sqrt{a^2x_0^2+4}\omega\cos\frac{2\omega\text{arccsch}(2/ax_0)}{a}+\frac{2+a^2x_0^2}{x_0}\sin\frac{2\omega\text{arccsch}(2/ax_0)}{a}\right]}{\sqrt{a^2x_0^2+4}x_0(a^2x_0^2+4)} \equiv F(\omega,x_0,a)\label{eqn:integral_k2KJ1}\\
  &\int_0^\infty k K_{i\omega/a}(k/a)^2dk=\frac{\pi\omega a}{2\sinh(\pi\omega/a)}\\
  &\int_0^\infty k^2 [K_{i\omega/a-1}(k/a)+K_{i\omega/a+1}(k/a)]K_{i\omega/a}(k/a)dk= \frac{\pi\omega a^2}{\sinh(\pi\omega/a)}
\end{align}

\section{Linear coupling}
\subsection{Calculation of $P$}
\begin{equation}
  \begin{split}
    P=&\lambda^2\int_0^\infty d\omega\iint d^2\mathbf{k}_{\bot}\left\{\exp\left[-\sigma^2(\Omega+\omega)^2\right]+\exp\left[-\sigma^2(\Omega-\omega)^2\right]e^{-2\pi\omega/a}\right\}\frac{\sinh(\pi\omega/a)}{2\pi^3a}\frac{K^2_{i\omega/a}\left(\frac{\kappa}{a}\right)}{1-e^{-2\pi\omega/a}}\\
    =&\frac{\lambda^2}{4\pi^3a}\int_0^\infty d\omega\iint d^2\mathbf{k}_{\bot}\left\{\exp\left[-\sigma^2(\Omega+\omega)^2\right]e^{\pi\omega/a}+\exp\left[-\sigma^2(\Omega-\omega)^2\right]e^{-\pi\omega/a}\right\}K^2_{i\omega/a}\left(\frac{\kappa}{a}\right)\\
    =&\frac{\lambda^2}{2\pi^2a}\int_0^\infty d\omega\left[e^{-\sigma^2(\Omega+\omega)^2+\pi\omega/a}+e^{-\sigma^2(\Omega-\omega)^2-\pi\omega/a}\right] \int_0^\infty dk_\bot k_\bot K^2_{i\omega/a}\left(\frac{\sqrt{k_\bot^2+m^2}}{a}\right)\\
    =&\frac{\lambda^2}{2\pi^2a}\int_0^\infty d\omega\left[e^{-\sigma^2(\Omega+\omega)^2+\pi\omega/a}+e^{-\sigma^2(\Omega-\omega)^2-\pi\omega/a}\right]
    \frac{m^2}{2}\left[K_{-1+i\omega/a}\left(\frac{m}{a}\right)K_{1+i\omega/a}\left(\frac{m}{a}\right)-K_{i\omega/a}\left(\frac{m}{a}\right)^2\right].{\tiny\text{(according to Mathematica)}}
  \end{split}
\end{equation}
The integral of $k_\bot$ can be done by
\begin{equation}
  \int_0^\infty dk_\bot k_\bot K^2_{i\omega/a}\left(\frac{k_\bot}{a}\right)
  =\frac{a\pi\omega}{2\sinh(\pi\omega/a)}.
\end{equation}
Thus,
\begin{equation}
  P=\frac{\lambda^2}{4\pi}\int_0^\infty d\omega \omega
  \frac{e^{-\sigma^2(\Omega+\omega)^2+\pi\omega/a}+e^{-\sigma^2(\Omega-\omega)^2-\pi\omega/a}}{\sinh(\pi\omega/a)}.
\end{equation}

\subsection{Calculation of $L_{AB}$}
The $\mu $s are given by
\begin{align}
  &\mu^A_{\omega\mathbf{k}_\bot}=\sqrt{2\pi}\exp\left[-\frac{\sigma_A^2}{2}(\Omega_A-\omega)^2\right]\left[\frac{\sinh(\pi\omega/a)}{4\pi^4a}\right]^{1/2}K_{i\omega/a}\left(\frac{\kappa}{a}\right),\\
  &\mu^B_{\omega\mathbf{k}_\bot}=\sqrt{2\pi}\exp\left[-\frac{\sigma_B^2}{2}(\Omega_B-\omega)^2+i(\Omega_B-\omega)\tau_0\right]\left[\frac{\sinh(\pi\omega/a)}{4\pi^4a}\right]^{1/2}K_{i\omega/a}\left(\frac{\kappa}{a}\right)e^{ik_xx_0},\\
  &\mu'^A_{\omega\mathbf{k}_\bot}=\sqrt{2\pi}\exp\left[-\frac{\sigma_A^2}{2}(\Omega_A+\omega)^2\right]\left[\frac{\sinh(\pi\omega/a)}{4\pi^4a}\right]^{1/2}K_{i\omega/a}\left(\frac{\kappa}{a}\right),\\
  &\mu'^B_{\omega\mathbf{k}_\bot}=\sqrt{2\pi}\exp\left[-\frac{\sigma_B^2}{2}(\Omega_B+\omega)^2+i(\Omega_B+\omega)\tau_0\right]\left[\frac{\sinh(\pi\omega/a)}{4\pi^4a}\right]^{1/2}K_{i\omega/a}\left(\frac{\kappa}{a}\right)e^{-ik_xx_0}.
\end{align}
Substituting these into \eqref{eqn: lwithv} we have
\begin{equation}
  \begin{split}
    L_{AB}=&\lambda_A\lambda_B\int_0^\infty d\omega\iint d^2\mathbf{k}_{\bot}\frac{\sinh(\pi\omega/a)}{2\pi^3a}\frac{K_{i\omega/a}\left(\frac{\kappa}{a}\right)^2}{1-e^{-2\pi\omega/a}}\\ &\times\left\{\exp\left[-\frac{\sigma_A^2}{2}(\Omega_A+\omega)^2-\frac{\sigma_B^2}{2}(\Omega_B+\omega)^2+i(\Omega_B+\omega)\tau_0-ik_xx_0\right]\right.\\ &\left.+\exp\left[-\frac{\sigma_A^2}{2}(\Omega_A-\omega)^2-\frac{\sigma_B^2}{2}(\Omega_B-\omega)^2+i(\Omega_B-\omega)\tau_0+ik_xx_0\right]e^{-2\pi\omega/a}\right\}\\
    =&\frac{\lambda_A\lambda_B}{4\pi^3a}\int_0^\infty d\omega\iint d^2\mathbf{k}_{\bot}K_{i\omega/a}\left(\frac{\kappa}{a}\right)^2\\ &\times\left\{\exp\left[-\frac{\sigma_A^2}{2}(\Omega_A+\omega)^2-\frac{\sigma_B^2}{2}(\Omega_B+\omega)^2+i(\Omega_B+\omega)\tau_0-ik_xx_0\right]e^{\pi\omega/a}\right.\\ &\left.+\exp\left[-\frac{\sigma_A^2}{2}(\Omega_A-\omega)^2-\frac{\sigma_B^2}{2}(\Omega_B-\omega)^2+i(\Omega_B-\omega)\tau_0+ik_xx_0\right]e^{-\pi\omega/a}\right\}\\
    &=\frac{\lambda_A\lambda_B}{4\pi^3a}\int_0^\infty d\omega\int_0^\infty k_\bot dk_\bot \int_0^{2\pi}d\alpha K_{i\omega/a}\left(\frac{\sqrt{k_\bot^2+m^2}}{a}\right)^2\\ &\times\left\{\exp\left[-\frac{\sigma_A^2}{2}(\Omega_A+\omega)^2-\frac{\sigma_B^2}{2}(\Omega_B+\omega)^2+i(\Omega_B+\omega)\tau_0-ik_\bot x_0\cos\alpha\right]e^{\pi\omega/a}\right.\\ &\left.+\exp\left[-\frac{\sigma_A^2}{2}(\Omega_A-\omega)^2-\frac{\sigma_B^2}{2}(\Omega_B-\omega)^2+i(\Omega_B-\omega)\tau_0+ik_\bot x_0\cos\alpha\right]e^{-\pi\omega/a}\right\}\\
    &=\frac{\lambda_A\lambda_B}{2\pi^2a}\int_0^\infty d\omega\int_0^\infty k_\bot dk_\bot K_{i\omega/a}\left(\frac{\sqrt{k_\bot^2+m^2}}{a}\right)^2J_0(k_\bot x_0)\\ &\times\left\{\exp\left[-\frac{\sigma_A^2}{2}(\Omega_A+\omega)^2-\frac{\sigma_B^2}{2}(\Omega_B+\omega)^2+i(\Omega_B+\omega)\tau_0\right]e^{\pi\omega/a}\right.\\ &\left.+\exp\left[-\frac{\sigma_A^2}{2}(\Omega_A-\omega)^2-\frac{\sigma_B^2}{2}(\Omega_B-\omega)^2+i(\Omega_B-\omega)\tau_0\right]e^{-\pi\omega/a}\right\}.
  \end{split}
\end{equation}
 We take the field to be massless and integrate over $k_\bot$ to obtain
\begin{equation}
\begin{split}
  L_{AB}=&\frac{\lambda_A\lambda_B}{2\pi^2a}\int_0^\infty d\omega \frac{\pi a}{x_0\sqrt{a^2x_0^2+4}\sinh(\pi\omega/a)}\sin\frac{2\omega\text{arcsinh}(ax_0/2)}{a}\\ &\times\left\{\exp\left[-\frac{\sigma_A^2}{2}(\Omega_A+\omega)^2-\frac{\sigma_B^2}{2}(\Omega_B+\omega)^2+i(\Omega_B+\omega)\tau_0\right]e^{\pi\omega/a}\right.\\ &\left.+\exp\left[-\frac{\sigma_A^2}{2}(\Omega_A-\omega)^2-\frac{\sigma_B^2}{2}(\Omega_B-\omega)^2+i(\Omega_B-\omega)\tau_0\right]e^{-\pi\omega/a}\right\}\\
  =&\frac{\lambda_A\lambda_B}{\pi}\frac{1}{x_0\sqrt{a^2x_0^2+4}}
  \int_0^\infty d\omega\sin\frac{2\omega\text{arcsinh}(ax_0/2)}{a} e^{-\sigma_A^2(\Omega_A^2+\omega^2)/2-\sigma_B^2((\Omega_B^2+\omega^2)/2+i\Omega_B\tau_0}\\ &\times\frac{\cosh\left(\sigma_A^2\Omega_A\omega+\sigma_B^2\Omega_B\omega-i\omega\tau_0-\pi\omega/a\right)}{\sinh(\pi\omega/a)}.
 \end{split}
\end{equation}

\subsection{Calculation of $M$}
We calculate the integrals \eqref{eqn: w1beforeana} and \eqref{eqn: w2beforeana} through analytic continuation
\begin{equation}
  \int_0^{\infty} du\frac{e^{-\sigma^2 u^2}}{u^2+a^2}=\frac{\pi e^{\sigma^2a^2}}{2a}\text{erfc}(\sigma a),
\end{equation}
where $\text{erfc}(z)$ is the complementary error function defined by
\begin{equation}
  \begin{split}
    \text{erfc}(z)=&1-\text{erf}(z)\\
    =&1-\frac{2}{\sqrt{\pi}}\int_0^z e^{-\eta^2}d\eta.
  \end{split}
\end{equation}
We have,
\begin{equation}
  \begin{split}
    \int_0^{\infty} du\frac{e^{-\sigma^2 u^2}}{u^2-\omega^2+ i\epsilon}
    =\frac{\pi e^{-\sigma^2\omega^2}}{2i\omega}\text{erfc}(i\sigma\omega),
  \end{split}
\end{equation}

\begin{equation}
  \begin{split}
 \int_0^{\infty} du\frac{e^{-\sigma^2 u^2}}{u^2-\omega^2- i\epsilon}
    =\frac{\pi e^{-\sigma^2\omega^2}}{2i\omega}\left[\text{erfc}(i\sigma\omega)-2\right].
\end{split}
\end{equation}

\section{Quadratic coupling}
\subsection{Calculation of $P$ and $L$}
We list the $\eta$s as follows.
\begin{align*}
  \eta^{0A}_{\omega_1\mathbf{k}_{1\bot},\omega_2\mathbf{k}_{2\bot}}=&\frac{\sqrt{2\pi}}{4\pi^4a} \exp\left[-\frac{\sigma_A^2}{2}(\omega_1+\omega_2+\Omega_A)^2\right] \sqrt{\sinh(\pi\omega_1/a)\sinh(\pi\omega_2/a)}K_{i\omega_1/a}\left(\frac{\kappa_1}{a}\right)K_{i\omega_2/a}\left(\frac{\kappa_2}{a}\right)\\
  \eta^{0B}_{\omega_1\mathbf{k}_{1\bot},\omega_2\mathbf{k}_{2\bot}}=&\frac{\sqrt{2\pi}}{4\pi^4a} \exp\left[-\frac{\sigma_B^2}{2}(\omega_1+\omega_2+\Omega_B)^2+i(\omega_1+\omega_2+\Omega_B)\tau_0\right] \sqrt{\sinh(\pi\omega_1/a)\sinh(\pi\omega_2/a)}\\
  &\times K_{i\omega_1/a}\left(\frac{\kappa_1}{a}\right)K_{i\omega_2/a}\left(\frac{\kappa_2}{a}\right) e^{-i(k_{1x}+k_{2x})x_0}\\
  \eta^{1A}_{\omega_1\mathbf{k}_{1\bot},\omega_2\mathbf{k}_{2\bot}}=&\frac{\sqrt{2\pi}}{4\pi^4a} \exp\left[-\frac{\sigma_A^2}{2}(\Omega_A-\omega_1+\omega_2)^2\right] \sqrt{\sinh(\pi\omega_1/a)\sinh(\pi\omega_2/a)}K_{i\omega_1/a}\left(\frac{\kappa_1}{a}\right)K_{i\omega_2/a}\left(\frac{\kappa_2}{a}\right)\\
  \eta^{1B}_{\omega_1\mathbf{k}_{1\bot},\omega_2\mathbf{k}_{2\bot}}=&\frac{\sqrt{2\pi}}{4\pi^4a} \exp\left[-\frac{\sigma_B^2}{2}(\Omega_B-\omega_1+\omega_2)^2+i(\Omega_B-\omega_1+\omega_2)\tau_0\right] \sqrt{\sinh(\pi\omega_1/a)\sinh(\pi\omega_2/a)}\\
  &\times K_{i\omega_1/a}\left(\frac{\kappa_1}{a}\right)K_{i\omega_2/a}\left(\frac{\kappa_2}{a}\right) e^{i(k_{1x}-k_{2x})x_0}\\
  \eta^{2A}_{\omega_1\mathbf{k}_{1\bot},\omega_2\mathbf{k}_{2\bot}}=&\frac{\sqrt{2\pi}}{4\pi^4a} \exp\left[-\frac{\sigma_A^2}{2}(\Omega_A+\omega_1-\omega_2)^2\right] \sqrt{\sinh(\pi\omega_1/a)\sinh(\pi\omega_2/a)}K_{i\omega_1/a}\left(\frac{\kappa_1}{a}\right)K_{i\omega_2/a}\left(\frac{\kappa_2}{a}\right)\\
  \eta^{2B}_{\omega_1\mathbf{k}_{1\bot},\omega_2\mathbf{k}_{2\bot}}=&\frac{\sqrt{2\pi}}{4\pi^4a} \exp\left[-\frac{\sigma_B^2}{2}(\Omega_B+\omega_1-\omega_2)^2+i(\Omega_B+\omega_1-\omega_2)\tau_0\right] \sqrt{\sinh(\pi\omega_1/a)\sinh(\pi\omega_2/a)}\\
  &\times K_{i\omega_1/a}\left(\frac{\kappa_1}{a}\right)K_{i\omega_2/a}\left(\frac{\kappa_2}{a}\right) e^{-i(k_{1x}-k_{2x})x_0}\\
  \eta^{12A}_{\omega_1\mathbf{k}_{1\bot},\omega_2\mathbf{k}_{2\bot}}=&\frac{\sqrt{2\pi}}{4\pi^4a} \exp\left[-\frac{\sigma_A^2}{2}(\Omega_A-\omega_1-\omega_2)^2\right] \sqrt{\sinh(\pi\omega_1/a)\sinh(\pi\omega_2/a)}K_{i\omega_1/a}\left(\frac{\kappa_1}{a}\right)K_{i\omega_2/a}\left(\frac{\kappa_2}{a}\right)\\
  \eta^{12B}_{\omega_1\mathbf{k}_{1\bot},\omega_2\mathbf{k}_{2\bot}}=&\frac{\sqrt{2\pi}}{4\pi^4a} \exp\left[-\frac{\sigma_B^2}{2}(\Omega_B-\omega_1-\omega_2)^2+i(\Omega_B-\omega_1-\omega_2)\tau_0\right] \sqrt{\sinh(\pi\omega_1/a)\sinh(\pi\omega_2/a)}.
\end{align*}
Substituting these to \eqref{eqn:2_quar_Pj0} and \eqref{eqn:2_quar_LAB0} and we get

\begin{equation}
  \begin{split}
    P_j=&\frac{2\lambda_j^2(2\pi)}{(4\pi^4a)^2}\int_0^\infty d\omega_1\int_0^\infty d\omega_2\iint d^2\mathbf{k}_{1\bot}
    \iint d^2\mathbf{k}_{2\bot} \frac{\sinh(\pi\omega_1/a)}{1-e^{-2\pi\omega_1/a}}\frac{\sinh(\pi\omega_2/a)}{1-e^{-2\pi\omega_2/a}}K_{i\omega_1/a}\left(\frac{\kappa_1}{a}\right)^2 K_{i\omega_2/a}\left(\frac{\kappa_2}{a}\right)^2\\
    &\times\left\{\exp\left[-\sigma_j^2(\omega_1+\omega_2+\Omega_j)^2\right] +\exp\left[-\sigma_j^2(\Omega_j+\omega_1-\omega_2)^2-2\pi\omega_2/a\right]\right.\\ &\left.+\exp\left[-\sigma_j^2(\Omega_j-\omega_1+\omega_2)^2-2\pi\omega_1/a\right]
    +\exp\left[-\sigma_j^2(\Omega_j-\omega_1-\omega_2)^2-2\pi(\omega_1+\omega_2)/a\right]\right\}\\
    =&\frac{2\lambda_j^2(2\pi)}{4(4\pi^4a)^2}\int_0^\infty d\omega_1\int_0^\infty d\omega_2\iint d^2\mathbf{k}_{1\bot}
    \iint d^2\mathbf{k}_{2\bot} K_{i\omega_1/a}\left(\frac{\kappa_1}{a}\right)^2 K_{i\omega_2/a}\left(\frac{\kappa_2}{a}\right)^2\\
    &\times\left\{\exp\left[-\sigma_j^2(\omega_1+\omega_2+\Omega_j)^2+\pi(\omega_1+\omega_2)/a\right] +\exp\left[-\sigma_j^2(\Omega_j+\omega_1-\omega_2)^2+\pi(\omega_1-\omega_2)/a\right]\right.\\ &\left.+\exp\left[-\sigma_j^2(\Omega_j-\omega_1+\omega_2)^2-\pi(\omega_1-\omega_2)/a\right]
    +\exp\left[-\sigma_j^2(\Omega_j-\omega_1-\omega_2)^2-\pi(\omega_1+\omega_2)/a\right]\right\}\\
    =&\frac{2\lambda_j^2(2\pi)^3}{4(4\pi^4a)^2}\int_0^\infty d\omega_1\int_0^\infty d\omega_2\int k_{1\bot}dk_{1\bot}\int k_{2\bot}dk_{2\bot} K_{i\omega_1/a}\left(\frac{\sqrt{k_{1\bot}^2+m^2}}{a}\right)^2 K_{i\omega_2/a}\left(\frac{\sqrt{k_{2\bot}^2+m^2}}{a}\right)^2\\
    &\times\left\{\exp\left[-\sigma_j^2(\omega_1+\omega_2+\Omega_j)^2+\pi(\omega_1+\omega_2)/a\right] +\exp\left[-\sigma_j^2(\Omega_j+\omega_1-\omega_2)^2+\pi(\omega_1-\omega_2)/a\right]\right.\\ &\left.+\exp\left[-\sigma_j^2(\Omega_j-\omega_1+\omega_2)^2-\pi(\omega_1-\omega_2)/a\right]
    +\exp\left[-\sigma_j^2(\Omega_j-\omega_1-\omega_2)^2-\pi(\omega_1+\omega_2)/a\right]\right\},
  \end{split}
\end{equation}
and
\begin{equation}
  \begin{split}
    L_{AB}=&\frac{2\lambda_A\lambda_B(2\pi)}{(4\pi^4a)^2}\int_0^\infty d\omega_1\int_0^\infty d\omega_2\iint d^2\mathbf{k}_{1\bot}
    \iint d^2\mathbf{k}_{2\bot} \frac{\sinh(\pi\omega_1/a)}{1-e^{-2\pi\omega_1/a}}\frac{\sinh(\pi\omega_2/a)}{1-e^{-2\pi\omega_2/a}}K_{i\omega_1/a}\left(\frac{\kappa_1}{a}\right)^2 K_{i\omega_2/a}\left(\frac{\kappa_2}{a}\right)^2\\
    &\times\left\{\exp\left[-\frac{\sigma_A^2}{2}(\omega_1+\omega_2+\Omega_A)^2-\frac{\sigma_B^2}{2}(\omega_1+\omega_2+\Omega_B)^2+i(\omega_1+\omega_2+\Omega_B)\tau_0-i(k_{1x}+k_{2x})x_0\right]\right.\\
    &+\exp\left[-\frac{\sigma_A^2}{2}(\Omega_A+\omega_1-\omega_2)^2-\frac{\sigma_B^2}{2}(\Omega_B+\omega_1-\omega_2)^2+i(\Omega_B+\omega_1-\omega_2)\tau_0-i(k_{1x}-k_{2x})x_0-2\pi\omega_2/a\right]\\
    &+\exp\left[-\frac{\sigma_A^2}{2}(\Omega_A-\omega_1+\omega_2)^2-\frac{\sigma_B^2}{2}(\Omega_B-\omega_1+\omega_2)^2+i(\Omega_B-\omega_1+\omega_2)\tau_0+i(k_{1x}-k_{2x})x_0-2\pi\omega_1/a\right]\\
    &\hspace{-1cm}\left.+\exp\left[-\frac{\sigma_A^2}{2}(\Omega_A-\omega_1-\omega_2)^2-\frac{\sigma_B^2}{2}(\Omega_B-\omega_1-\omega_2)^2+i(\Omega_B-\omega_1-\omega_2)\tau_0+i(k_{1x}+k_{2x})x_0-2\pi(\omega_1+\omega_2)/a\right]\right\}\\
    &\hspace{-3cm}=\frac{2\lambda_A\lambda_B(2\pi)^3}{4(4\pi^4a)^2}\int_0^\infty d\omega_1\int_0^\infty d\omega_2\int k_{1\bot}dk_{1\bot}\int k_{2\bot}dk_{2\bot} K_{i\omega_1/a}\left(\frac{\sqrt{k_{1\bot}^2+m^2}}{a}\right)^2 K_{i\omega_2/a}\left(\frac{\sqrt{k_{2\bot}^2+m^2}}{a}\right)^2J_0(k_{1\bot}x_0)J_0(k_{2\bot}x_0)\\
    &\hspace{-1cm}\times\left\{\exp\left[-\frac{\sigma_A^2}{2}(\omega_1+\omega_2+\Omega_A)^2-\frac{\sigma_B^2}{2}(\omega_1+\omega_2+\Omega_B)^2+i(\omega_1+\omega_2+\Omega_B)\tau_0+\pi(\omega_1+\omega_2)/a\right]\right.\\
    &\hspace{-1cm}+\exp\left[-\frac{\sigma_A^2}{2}(\Omega_A+\omega_1-\omega_2)^2-\frac{\sigma_B^2}{2}(\Omega_B+\omega_1-\omega_2)^2+i(\Omega_B+\omega_1-\omega_2)\tau_0+\pi(\omega_1-\omega_2)/a\right]\\
    &\hspace{-1cm}+\exp\left[-\frac{\sigma_A^2}{2}(\Omega_A-\omega_1+\omega_2)^2-\frac{\sigma_B^2}{2}(\Omega_B-\omega_1+\omega_2)^2+i(\Omega_B-\omega_1+\omega_2)\tau_0-\pi(\omega_1-\omega_2)/a\right]\\
    &\hspace{-1cm}\left.+\exp\left[-\frac{\sigma_A^2}{2}(\Omega_A-\omega_1-\omega_2)^2-\frac{\sigma_B^2}{2}(\Omega_B-\omega_1-\omega_2)^2+i(\Omega_B-\omega_1-\omega_2)\tau_0-\pi(\omega_1+\omega_2)/a\right]\right\}.
  \end{split}
\end{equation}

\subsection{Calculation of $M$}
The $W$s are listed as follows.
\begin{align*}
W^0_{\omega_1\mathbf{k}_{1\bot},\omega_2\mathbf{k}_{2\bot}}=&\frac{\sinh(\pi\omega_1/a)\sinh(\pi\omega_2/a)}{(4\pi^4a)^2/(2\pi)}\frac{-2(\omega_1+\omega_2)}{2\pi i}K_{i\omega_1/a}\left(\frac{\kappa_1}{a}\right)^2K_{i\omega_2/a}\left(\frac{\kappa_2}{a}\right)^2\\
  &\times\int du \frac{\exp\left[-\frac{\sigma_A^2}{2}(\Omega_A+u)^2\right]\exp\left[-\frac{\sigma_B^2}{2}(\Omega_B-u)^2+i(\Omega_B-u)\tau_0\right]}{u^2-(\omega_1+\omega_2)^2+i\epsilon}
  e^{i(k_{1x}+k_{2x})x_0}\\
W^1_{\omega_1\mathbf{k}_{1\bot},\omega_2\mathbf{k}_{2\bot}}=&\frac{\sinh(\pi\omega_1/a)\sinh(\pi\omega_2/a)}{(4\pi^4a)^2/(2\pi)}\frac{2(\omega_1-\omega_2)}{2\pi i} K_{i\omega_1/a}\left(\frac{\kappa_1}{a}\right)^2K_{i\omega_2/a}\left(\frac{\kappa_2}{a}\right)^2 \\
    &\times\int du \frac{\exp\left[-\frac{\sigma_A^2}{2}(\Omega_A+u)^2\right]\exp\left[-\frac{\sigma_B^2}{2}(\Omega_B-u)^2+i(\Omega_B-u)\tau_0\right]}{u^2-(\omega_1-\omega_2)^2-i\epsilon(\omega_1-\omega_2)}
    e^{i(k_{1x}-k_{2x})x_0}\\
W^2_{\omega_1\mathbf{k}_{1\bot},\omega_2\mathbf{k}_{2\bot}}=&\frac{\sinh(\pi\omega_1/a)\sinh(\pi\omega_2/a)}{(4\pi^4a)^2/(2\pi)}\frac{-2(\omega_1-\omega_2)}{2\pi i}K_{i\omega_1/a}\left(\frac{\kappa_1}{a}\right)^2K_{i\omega_2/a}\left(\frac{\kappa_2}{a}\right)^2\\
&\times \int du \frac{\exp\left[-\frac{\sigma_A^2}{2}(\Omega_A+u)^2\right]\exp\left[-\frac{\sigma_B^2}{2}(\Omega_B-u)^2+i(\Omega_B-u)\tau_0\right]}{u^2-(\omega_1-\omega_2)^2+i\epsilon(\omega_1-\omega_2)}
    e^{i(k_{1x}-k_{2x})x_0}\\
W^{12}_{\omega_1\mathbf{k}_{1\bot},\omega_2\mathbf{k}_{2\bot}}=&\frac{\sinh(\pi\omega_1/a)\sinh(\pi\omega_2/a)}{(4\pi^4a)^2/(2\pi)}\frac{2(\omega_1+\omega_2)}{2\pi i}K_{i\omega_1/a}\left(\frac{\kappa_1}{a}\right)^2 K_{i\omega_2/a}\left(\frac{\kappa_2}{a}\right)^2\\
    &\times\int du\frac{\exp\left[-\frac{\sigma_A^2}{2}(\Omega_A+u)^2\right]\exp\left[-\frac{\sigma_B^2}{2}(\Omega_B-u)^2+i(\Omega_B-u)\tau_0\right]}{u^2-(\omega_1+\omega_2)^2-i\epsilon}
     e^{i(k_{1x}+k_{2x})x_0}.
\end{align*}
The integral over u can be done using the method of analytic continuation as in section 3 wiht ($\sigma_A=\sigma_B=\sigma,\Omega_A=\Omega_B=\Omega,\tau_0=0$).
\begin{align}
  \begin{split}
    W^0_{\omega_1\mathbf{k}_{1\bot},\omega_2\mathbf{k}_{2\bot}}=&\frac{\sinh(\pi\omega_1/a)\sinh(\pi\omega_2/a)}{(4\pi^4a)^2/(2\pi)}\frac{-4(\omega_1+\omega_2)}{2\pi i}K_{i\omega_1/a}\left(\frac{\kappa_1}{a}\right)^2K_{i\omega_2/a}\left(\frac{\kappa_2}{a}\right)^2e^{-\sigma^2\Omega^2}\\
  &\times e^{i(k_{1x}+k_{2x})x_0}\int_0^\infty du \frac{e^{-\sigma^2u^2}}{u^2-(\omega_1+\omega_2)^2+i\epsilon}\\
  =&\frac{\sinh(\pi\omega_1/a)\sinh(\pi\omega_2/a)}{(4\pi^4a)^2/(2\pi)} K_{i\omega_1/a}\left(\frac{\kappa_1}{a}\right)^2K_{i\omega_2/a}\left(\frac{\kappa_2}{a}\right)^2e^{-\sigma^2\Omega^2}\\
  &\times e^{i(k_{1x}+k_{2x})x_0}e^{-\sigma^2(\omega_1+\omega_2)^2}\text{erfc}[i\sigma(\omega_1+\omega_2)]
  \end{split}\\
  \begin{split}
    W^1_{\omega_1\mathbf{k}_{1\bot},\omega_2\mathbf{k}_{2\bot}}=&\frac{\sinh(\pi\omega_1/a)\sinh(\pi\omega_2/a)}{(4\pi^4a)^2/(2\pi)}\frac{4(\omega_1-\omega_2)}{2\pi i} K_{i\omega_1/a}\left(\frac{\kappa_1}{a}\right)^2K_{i\omega_2/a}\left(\frac{\kappa_2}{a}\right)^2 e^{-\sigma^2\Omega^2}\\
    &\times e^{i(k_{1x}-k_{2x})x_0}\int_0^\infty du \frac{e^{-\sigma^2u^2}}{u^2-(\omega_1-\omega_2)^2-i\epsilon(\omega_1-\omega_2)}\\
    =&-\frac{\sinh(\pi\omega_1/a)\sinh(\pi\omega_2/a)}{(4\pi^4a)^2/(2\pi)}K_{i\omega_1/a}\left(\frac{\kappa_1}{a}\right)^2K_{i\omega_2/a}\left(\frac{\kappa_2}{a}\right)^2 e^{-\sigma^2\Omega^2}\\
    &\times e^{i(k_{1x}-k_{2x})x_0}e^{-\sigma^2(\omega_1-\omega_2)^2}\Big[\text{erfc}[i\sigma(\omega_1-\omega_2)]-1-\text{sgn}(\omega_1-\omega_2)\Big]
  \end{split}\\
  \begin{split}
    W^2_{\omega_1\mathbf{k}_{1\bot},\omega_2\mathbf{k}_{2\bot}}=&\frac{\sinh(\pi\omega_1/a)\sinh(\pi\omega_2/a)}{(4\pi^4a)^2/(2\pi)}\frac{-4(\omega_1-\omega_2)}{2\pi i}e^{-\sigma^2\Omega^2}K_{i\omega_1/a}\left(\frac{\kappa_1}{a}\right)^2K_{i\omega_2/a}\left(\frac{\kappa_2}{a}\right)^2\\
    &\times e^{i(k_{1x}-k_{2x})x_0} \int_0^\infty du \frac{e^{-\sigma^2u^2}}{u^2-(\omega_1-\omega_2)^2+i\epsilon(\omega_1-\omega_2)}\\
    =&\frac{\sinh(\pi\omega_1/a)\sinh(\pi\omega_2/a)}{(4\pi^4a)^2/(2\pi)}e^{-\sigma^2\Omega^2}K_{i\omega_1/a}\left(\frac{\kappa_1}{a}\right)^2K_{i\omega_2/a}\left(\frac{\kappa_2}{a}\right)^2\\
    &\times e^{i(k_{1x}-k_{2x})x_0} e^{-\sigma^2(\omega_1-\omega_2)^2}\Big[\text{erfc}[i\sigma(\omega_1-\omega_2)]-1+\text{sgn}(\omega_1-\omega_2)\Big]
  \end{split}\\
  \begin{split}
    W^{12}_{\omega_1\mathbf{k}_{1\bot},\omega_2\mathbf{k}_{2\bot}}=&\frac{\sinh(\pi\omega_1/a)\sinh(\pi\omega_2/a)}{(4\pi^4a)^2/(2\pi)}\frac{4(\omega_1+\omega_2)}{2\pi i}e^{-\sigma^2\Omega^2}K_{i\omega_1/a}\left(\frac{\kappa_1}{a}\right)^2 K_{i\omega_2/a}\left(\frac{\kappa_2}{a}\right)^2\\
    &\times e^{i(k_{1x}+k_{2x})x_0}\int_0^\infty du\frac{e^{-\sigma^2u^2}}{u^2-(\omega_1+\omega_2)^2-i\epsilon}\\
     =&-\frac{\sinh(\pi\omega_1/a)\sinh(\pi\omega_2/a)}{(4\pi^4a)^2/(2\pi)}e^{-\sigma^2\Omega^2}K_{i\omega_1/a}\left(\frac{\kappa_1}{a}\right)^2 K_{i\omega_2/a}\left(\frac{\kappa_2}{a}\right)^2\\
    &\times e^{i(k_{1x}+k_{2x})x_0}e^{-\sigma^2(\omega_1+\omega_2)^2}\Big[\text{erfc}[i\sigma(\omega_1+\omega_2)]-2\Big]
  \end{split}
\end{align}


\begin{thebibliography}{99}

\bibitem{Wald-1995}
Robert M. Wald,
\newblock {\em Quantum field theory in curved spacetime and black hole thermodynamics},
\newblock University of Chicago Press, Chicago (1994).

\bibitem{Unruh-1976}
W. G. Unruh,
\newblock ``Notes on black hole evaporation,''
\newblock Phys. Rev. D {\bf 14}, 870 (1976).

\bibitem{Hawking-1975}
S. W. Hawking,
\newblock ``Particle Creation by Black Holes,''
\newblock Commun. Math. Phys. {\bf 43}, 199 (1975) [Erratum-ibid. 46, 206 (1976)].

\bibitem{DeWitt-1979}
B. DeWitt,
\newblock {\em General Relativity: An Einstein Centenary Survey}, edited by S.W. Hawking and W. Israel
\newblock (Cambridge University Press, Cambridge, 1979).

\bibitem{Unruh-1984}
W. G. Unruh and R. M.Wald,
\newblock ``What happens when an accelerating observer detects a Rindler particle,''
\newblock Phys. Rev. D {\bf 29}, 1047 (1984).

\bibitem{Pozas-Kerstjens-2016}
A. Pozas-Kerstjens and E. Martín-Martínez,
\newblock ``Entanglement harvesting from the electromagnetic vacuum with hydrogenlike atoms,''
\newblock Phys. Rev. D {\bf 94}, 064074.

\bibitem{Takagi-1986}
S. Takagi,
\newblock ``Vacuum Noise and Stress Induced by Uniform Acceleration,''
\newblock Prog. Theor. Phys. Suppl. {\bf 88}, 1 (1986)

\bibitem{Sachs-2017}
Allison Sachs, Robert B. Mann, and Eduardo Martín-Martínez,
\newblock ``Entanglement harvesting and divergences in quadratic Unruh-DeWitt detector pairs,''
\newblock Phys. Rev. D {\bf 96}, 085012.

\bibitem{Reeh-1961}
H. Reeh and S. Schlieder,
\newblock ``Bemerkungen zur unitäräquivalenz von lorentzinvarianten feldern,''
\newblock Nuovo Cimento {\bf 22} (1961) 1051.

\bibitem{Maldacena-2003}
Juan Maldacena,
\newblock ``Eternal black holes in anti-de Sitter,''
\newblock JHEP {\bf 04} (2003) 021.

\bibitem{Salton-2015}
Grant Salton, Robert B Mann, and Nicolas C Menicucci,
\newblock ``Acceleration-assisted entanglement harvesting and rangefinding,''
\newblock New J. Phys. {\bf 17} 035001.

\bibitem{Liu-2021}
Zhihong Liu, Jialin Zhang, Robert B. Mann, and Hongwei Yu,
\newblock ``Does acceleration assist entanglement harvesting?''
\newblock 	arXiv:2111.04392.


\bibitem{Yu-2009}
Ting Yu and J. H. Eberly,
\newblock ``Sudden Death of Entanglement,''
\newblock Science {\bf 323}(5914), pp.598-601,2009.


\bibitem{Crispino-2008}
L. C. B. Crispino, A. Higuchi and G. E. A. Matsas,
\newblock ``The Unruh effect and its applications,''
\newblock Rev. Mod. Phys. {\bf 80}, 787 (2008).	

\bibitem{Hummer-2016}
Daniel Hümmer, Eduardo Martín-Martínez and Achim Kempf,
\newblock ``Renormalized Unruh-DeWitt particle detector models for boson and fermion fields,''
\newblock Phys. Rev. D {\bf 93}, 024019 (2016).

\end{thebibliography}
\end{document}